\documentclass[12pt,preprint]{aastex}

\shorttitle{Generic Aberration Patterns}
\shortauthors{Schechter and Levinson}
\usepackage{graphicx}    
\usepackage{amsmath}   
\usepackage{lscape}
\usepackage{multirow}
\usepackage{natbib}
\usepackage{afterpage}
\usepackage[bottom]{footmisc}

\begin{document}

\title{Generic Misalignment Aberration Patterns in Wide-Field Telescopes}

\author{Paul L. Schechter}
\affil{MIT Kavli Institute, Cambridge, MA 02139}
\email{schech@mit.edu}
\and
\author {Rebecca Sobel Levinson}
\affil{MIT Kavli Institute, Cambridge, MA 02139}
\email{rsobel@mit.edu}

\begin{abstract}

Axially symmetric telescopes produce well known ``Seidel'' off-axis
third-order aberration patterns: coma, astigmatism, curvature of field
and distortion.  When axial symmetry is broken by the small
misalignments of optical elements, additional third-order aberration
patterns arise: one each for coma, astigmatism and curvature of field
and two for distortion.  Each of these misalignment patterns is
characterized by an associated two-dimensional vector, each of which
in turn is a linear combination of the tilt and decenter vectors of
the individual optical elements.  For an $N$-mirror telescope,
$2(N-1)$ patterns must be measured to keep the telescope aligned.  
Alignment of the focal plane may require two additional patterns.  For
$N=3$, as in a three mirror anastigmat, there is a two-dimensional
``subspace of benign misalignment'' over which the misalignment
patterns for third-order coma, astigmatism and curvature of field are
identically zero.  One would need to measure at least one of the two
distortion patterns to keep the telescope aligned.  Alternatively, one
might measure one of the fifth-order misalignment patterns, which are
derived herein.  But the fifth-order patterns are rather
insensitive to misalignments, even with moderately wide fields,
rendering them of relatively little use in telescope alignment.
Another alternative would be to use telescope pointing as part of the
alignment solution.

\end{abstract}

\section{Introduction}
\label{sec:introduction}

\subsection{new telescopes, stringent constraints}

The designs for several large telescopes that may get built in the
next decade are driven largely by the need for superb and stable image
quality over wide fields \citep{MaBernstein2008, 
PhillionOliver2006}.  Perhaps the most
  demanding of the scientific programs that in turn drive these
  requirements is that of cosmological weak lensing, using galaxy
  images to measure gravitational shape distortions as small as as one
  part in ten thousand.

Of particular concern for ground-based telescopes are the rigid body
motions of the optical elements due to gravitational and thermal
stresses on the telescope structure.\footnote{Such telescopes are 
{\it also} subject to deformations of the optical elements, but on
  longer timescales than the rigid body motions.}  The positions of
the optical elements can be controlled only to the extent that they
can be measured, putting a premium on the accurate characterization of
the aberrations generated by telescope misalignments.

On the assumption that at least some of these aberrations are best
measured with wavefront sensors, several questions immediately arise.
How many aberrations must be measured?  Which ones?  At how many field
positions must the measurements be made?  Where?

One can always, as a last resort, attempt to answer these questions by
simulation.  But we argue here that the answers to these questions
hinge on the identification of generic field aberration patterns that
emerge in a wide variety of circumstances.  A relatively small number
of such patterns suffices to efficiently diagnose and correct
misalignments.  And only a small number of field points must be
sampled to measure these patterns.  Moreover two of the patterns 
may in some cases be determined directly from
science data.

\subsection{literature} 

The practical astronomical literature on the alignment of wide field
telescopes is very limited.  \citeauthor{McLeod1996}'s
\citeyearpar{McLeod1996} paper describing the use of coma and
astigmatism patterns to align the Whipple Observatory 1.2-m telescope
anchors the recent literature.  \citet{WilsonDelabre1997} discuss the
alignment of the ESO NTT.  \citet{GittonNoethe1998} describe the
alignment of the ESO VLTs, and \citet{NoetheGuisard2000} give a more
general description of the astigmatism patterns expected from
two-mirror telescopes.  \citet{LeeDaltonToshKim2008} give a general
treatment of third-order misalignment distortions and then discuss
coma and astigmatism and curvature of field in their case studies.
\citet{PalunasFloyd2010} describe the alignment of the Magellan
Nasmyth telescopes using coma, astigmatism and curvature of field.

\citet{Marechal1950} derives the third-order misalignment aberration
patterns for coma, astigmatism, curvature of field and distortion.

Thompson and collaborators \citep{ShackThompson1980, Thompson2005,
ThompsonSchmid2009} develop a formalism for analyzing telescope
misalignments using a vector notation that is elegant and relatively
transparent.  It isolates generic misalignment patterns associated
with third-order aberrations -- coma, astigmatism, curvature of field
and distortion -- and beyond that, generic misalignment patterns
associated with fifth-order aberrations.  

In an unpublished M.S. thesis, \citet{Tessieres2003} used ray tracing
software to determine amplitudes for Thompson's misalignment patterns,
which at that time had only appeared in Thompson's Ph.D. thesis
\citeyearpar{Thompson1980}.  \citet{HviscBurge2008} build on
Tessieres' work in modeling a four mirror corrector for the Hobby-Ebberly
Telescope.  They identify the linear combinations of orthogonal
aberration patterns (integrated over the field) that are most
sensitive to the tilts and decenters of the mirrors.

\subsection{outline}

In the following sections we rederive these same misalignment
aberration patterns, using the Thompson et al. vector notation but
following instead the development of \citet{Schroeder1987}.  The present
paper is quite similar in spirit to \citeauthor{Tessieres2003}': identify those
aberration patterns of potential interest and ascertain which are of
greatest value in aligning a telescope.  However our approach differs
in that, in the interest of efficiency, we ignore patterns that are
non-linear in the tilts and decenters of the mirrors.  This
simplification is appropriate for small misalignments of an otherwise
rotationally symmetric telescope.  We also give greater
consideration to the role of distortion, curvature of field and
spherical aberration than did Tessieres.

In \S \ref{sec:generic_patterns} we discuss the misalignment patterns
produced by two-mirror telescopes, proceeding from the better known
generic coma and astigmatism misalignment patterns, through the almost
trivial curvature of field misalignment pattern, to the two distortion
misalignment patterns.  We then retrace our steps using a more general
approach that shows how the misalignment patterns produced by an optic
derive from the surface of that optic.  In \S
\ref{sec:aligning_3-mirror} we discuss the alignment of three-mirror
telescopes.  

In \S \ref{sec:generic_fifth-order} we use the
same methods used in \S \ref{sec:generic_patterns} to deduce the
misalignment patterns associated with fifth-order aberrations.  In the
course of this we attempt to systematize the somewhat ragged
nomenclature associated with the fifth-order aberrations.  We additionally 
examine the relative magnitudes of the fifth-order aberrations, which cast 
some doubt on their practical utility for telescope alignment.  In \S
\ref{sec:WFS} we discuss the number and placement of wavefront 
sensors needed to align a three mirror telescope. In \S
\ref{sec:complications} we address a variety of complicating factors:
mirror deformations, transmitting correctors, central obscurations and
focal plane tilts.  In \S \ref{sec:to_use_or_not_to_use} we discuss
several ways in which the misalignment aberration patterns might be
used, and why one might ultimately choose to forego their use.

\section{Generic patterns and two-mirror telescopes}
\label{sec:generic_patterns}

\subsection{coma} 

\citeauthor{McLeod1996}'s \citeyearpar{McLeod1996} paper shows how an
astigmatic misalignment pattern can be used in conjunction with coma to
align a telescope.  \citeauthor{McLeod1996}'s first step is to center the secondary so
as to zero the coma.  He does not explicitly refer to a coma
``pattern,'' but it is widely appreciated that decentering the
secondary of a Ritchey-Chretien telescope produces coma that is to
first order constant across the field.  For the present purposes we
take this to be a pattern, albeit a boring one.  \citeauthor{McLeod1996} does identify
an astigmatism pattern, which he then renders symmetric by rotating the
secondary about its coma-free pivot.

\citet{Schroeder1987} calculates the coma patterns that arise in a two
mirror telescope, allowing for tilting and decentering the secondary.
The comatic wavefront $G^{coma}$ is a function of position on the
pupil $\vec \rho$, with polar coordinates $\rho$ and $\phi$, and 
field angle $\vec \sigma$, with polar coordinates $\sigma$ and $\theta$: 

\begin{equation}
\label{eq:coma_s2} 
G^{coma} = G^{coma}_{Seidel}     \sigma  \rho^3\cos(\phi - \theta)
        +  G^{coma}_{decenter}   \ell   \rho^3\cos(\phi - \phi_\ell)
        +  G^{coma}_{tilt}      \alpha  \rho^3\cos(\phi - \phi_\alpha)
\end{equation}

\noindent 
where $\vec \ell$ is the decenter of the secondary which projects to
angle $\phi_\ell$ on the pupil, and $\vec \alpha$ is the vector tilt of
the secondary which projects to angle $\phi_\alpha$ on the pupil.  The
first term on the right hand side gives the symmetric ``Seidel'' coma
typical of an aligned two mirror telescope.  The next two terms give the
constant coma pattern typical of a decentering or tilt of the
secondary.  The coefficients $G^{coma}_{Seidel}$,
$G^{coma}_{decenter}$, and $G^{coma}_{tilt}$ depend upon the radii of
curvature, $R_i$,  and conic constants, $K_i$, of the two mirrors, the
indices of refraction of the material (air) preceding the mirrors, $n_i$, the
positions, $s_i$, and magnifications, $m_i$ of the object for each mirror, 
and the distance from the primary mirror to the secondary, $W$.
In a Ritchey-Chretien telescope the $G^{coma}_{Seidel}$ term is identically
zero, giving no coma when aligned.  The notation here is different
from that of \citeauthor{Schroeder1987}; the conversion from
\citeauthor{Schroeder1987}'s to the present notation is given in
appendix A.

\begin{figure}[htb]
\vspace{3.2 truein} \includegraphics{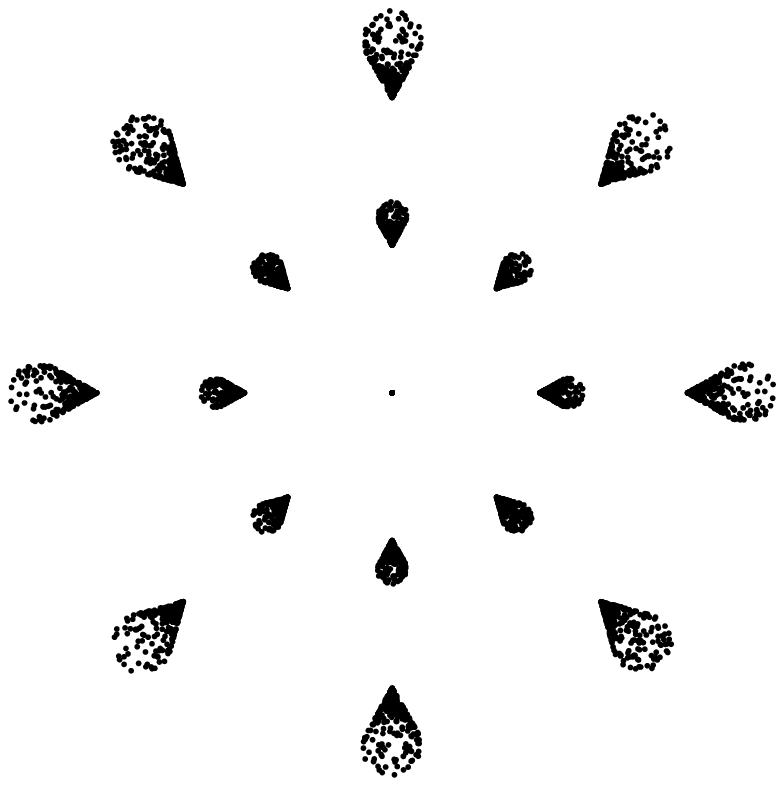}
\includegraphics{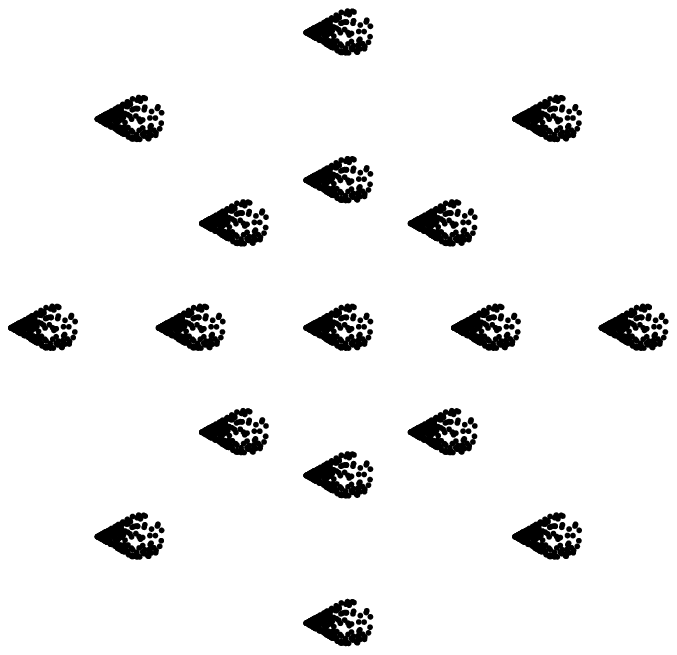} \figcaption{Comatic field patterns.  (a) 
  Seidel coma field pattern typical of an aligned telescope.
  (b) Constant coma indicating a
  tilt and (or) decentering of
  the secondary along the x axis.\label{fig:coma_s2} }
\end{figure}

The Seidel term varies linearly as field angle $\sigma$ and as
$\rho^3\cos\phi$ and (or) $\rho^3\sin\phi$ on the pupil.   The
tilt and decenter terms are constant over the field, but have the same
functional dependence on pupil coordinates as the first term.  We
shall somewhat loosely refer to any aberration that has the same
dependence upon pupil coordinates as ``coma,'' even when it does not
have the Seidel coma dependence on field angle $\sigma$.  Figure\ \ref{fig:coma_s2}a 
shows the point spread function at various points in the field for the first
term in equation \eqref{eq:coma_s2}.  Figure\ \ref{fig:coma_s2}b shows the point 
spread function pattern typical of either of the last two terms in equation \eqref
{eq:coma_s2}.

In practice, the constant coma pattern shown in Figure\ \ref{fig:coma_s2}b 
will be superimposed on the Seidel pattern if one is present.  We here plot the two 
patterns separately as the patterns have different physical motivations and thus 
provide different information about the telescope.  The Seidel pattern is a result 
of the telescope design, and fitting for that pattern provides no information about tilts or decenterings 
of the mirrors.  In contrast, the magnitude and orientation of the 
constant coma pattern {\it will} provide information about a telescope's alignment.

\subsection{astigmatism}

Following Schroeder's (1987) example for coma,
\citet{McLeod1996} calculated the corresponding astigmatism pattern for the
case of a nulled field constant coma pattern.  He finds an astigmatic
wavefront, $G^{astig}$, given by:

\begin{equation}
\label{eq:astig_s2}
G^{astig} = G^{astig}_{sym}     \sigma^2 \rho^2\cos2(\phi - \theta)
        +  G^{astig}_{decenter}   \sigma\ell   \rho^2\cos(2\phi - \theta - \phi_\ell)
        +  G^{astig}_{tilt}       \sigma\alpha  \rho^2\cos(2\phi - \theta - \phi_\alpha)
\end {equation}

\noindent
The first term on the right gives the symmetric astigmatism typical of
an aligned two mirror telescope.  The next two terms give the
astigmatism pattern typical of a decentering or tilt of the secondary.
The coefficients $G^{astig}_{sym}$, $G^{astig}_{decenter}$ and
$G^{astig}_{tilt}$ again depend upon the radii of curvature, $R_i$ and
conic constants, $K_i$ of the two mirrors, the indices of refraction
of the air preceding the mirrors, $n_i$, the
positions, $s_i$ and
magnifications, $m_i$ of the object for each mirror, and the distance
from the primary mirror to the secondary, $W$.  The details of the
conversion from \citeauthor{McLeod1996}'s notation to the above are
given in appendix\ B.

\begin{figure}[htb]
\vspace{3.0 truein}
\includegraphics{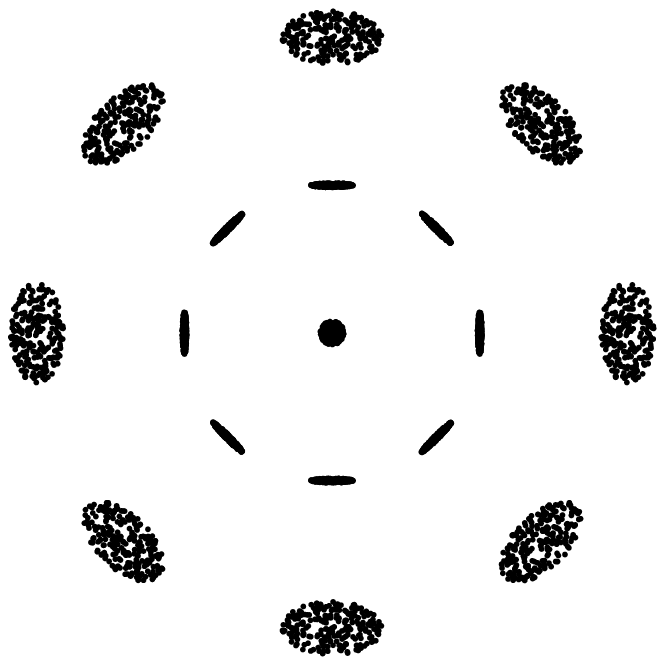}
\includegraphics{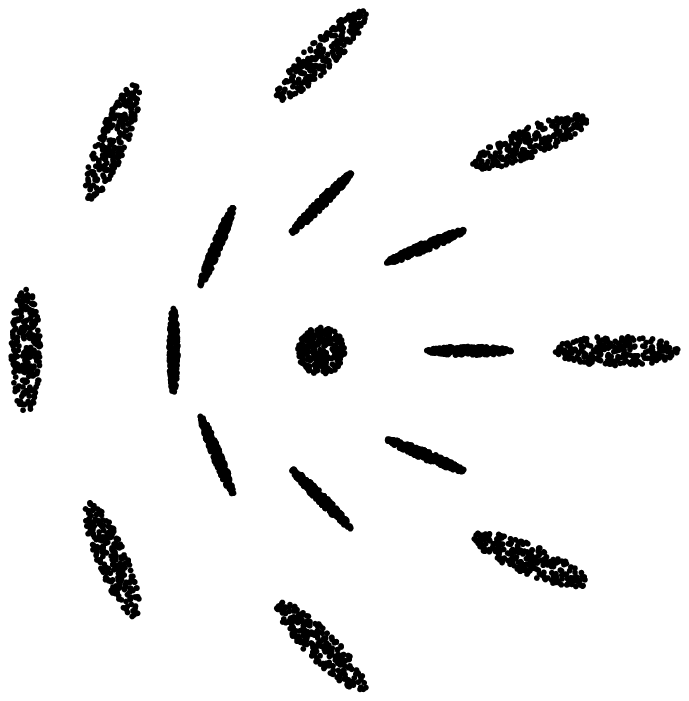}
\caption{Astigmatic field patterns.  (a) Seidel astigmatism field pattern
  typical of an aligned telescope.  A constant defocus has been
  added to show the orientation of the astigmatism.
  (b) Astigmatic field pattern
  indicating a tilt and (or)
  decentering of the secondary along the x axis.\label{fig:astig_s2} }
\end{figure}

The symmetric term varies as the square of the field radius $\sigma$
and varies as $\rho^2\cos2\phi$ and (or) $\rho^2\sin2\phi$ on the
pupil.  This is almost, but not quite the variation associated with
third-order, Seidel astigmatism.\footnote {By convention ``Seidel''
  astigmatism is taken to vary as $\rho^2\cos^2\phi$ on the pupil.  By
  contrast ``Zernike'' astigmatism is almost always taken to vary as
  $\rho^2 \cos 2\phi.$ The Seidel definition emerges naturally from
  the derivation of aberration patterns.  The Zernike definition makes
  for more symmetric wavefronts and orthogonality among the different
  aberrations.  A similar ambiguity arises in the definition of
  trefoil.}  It is readily decomposed into terms that vary as
$\rho^2\cos^2\phi$ (Seidel astigmatism) and $\rho^2\cos^0\phi$ (Seidel
curvature of field).  The tilt and decenter terms have the same
functional dependence on pupil coordinates as the symmetric term, but
vary linearly with distance from the center of the field $\sigma$, and
vary as the cosine and (or) sine of the field angle $\theta$.  We
shall again refer loosely to any aberration that has the same
dependence upon pupil coordinates as astigmatism, even when it does
not have the Seidel dependence on field position.
Figure\ \ref{fig:astig_s2}a shows the point spread function at various
points in the field for the symmetric term in equation
\eqref{eq:astig_s2}.  Figure\ \ref{fig:astig_s2}b shows the point
spread function pattern characteristic of either the tilt or decenter
terms in equation \eqref{eq:astig_s2}.\footnote{The ``dreamcatcher'' plot 
of Figure 2b makes cameo
appearances in a number of contexts.  It can be seen in a map of image
elongations at the prime focus of the LBT
\citep{Romano2010} and in the point spread function map, Figure 4.14, in
version 12.0 of the Chandra Proposer's Observatory Guide 
\citep{Chandra2009}.  The first such plot of which the authors are
aware is in the paper by \citet{ShackThompson1980}.  \citet{Marechal1950}
comes close plotting the magnitude and orientation of
the misalignment astigmatism pattern but suppressing the sign.}

As with the comatic aberrations,  the astigmatic field pattern
shown in Figure\ \ref{fig:astig_s2}b will in general be superimposed on the Seidel 
pattern if one is present.  Some previous treatments have referred to 
this superposition of field patterns as a single misalignment pattern;
\citet{McLeod1996} describes a single astigmatism pattern, which he
symmetrizes in the course of aligning his telescope; \citet{Thompson2005}
refers to a binodal astigmatism pattern which results from misalignments.
In this paper we decompose the astigmatism into two patterns, a symmetric 
one characteristic of an aligned telescope and an asymmetric one 
introduced by misalignments.  The asymmetric pattern described by 
\citet{McLeod1996} and the binodal pattern plotted by \citet{Thompson2005} 
are produced by superimposing the symmetric and asymmetric patterns in 
Figures \ \ref{fig:astig_s2}a and \ \ref{fig:astig_s2}b.  The nodes are simply field 
positions where the symmetric Seidel and asymmetric misalignment patterns 
have cancelled.\footnote{Terms that are nonlinear in the misalignments 
also contribute to the positions of the nodes.  Most interestingly, the inclusion
of a non-linear term may rotate the orientation of the nodes by $90^{\circ}$.
However, the inclusion of terms that are nonlinear in the misalignment has 
no effect on the underlying Seidel or linear patterns. }

Both for coma and for astigmatism (and in the cases of the additional
aberrations considered below) the misalignment pattern varies as one
power of field radius less rapidly than the corresponding symmetric
pattern.  

\subsection{curvature of field}

Coma and astigmatism are just two of the five third-order Seidel
aberrations.  Curvature of field (henceforth COF) manifests itself as
a defocus that varies as the square of the distance from the center
of an assumed flat focal plane.  \citeauthor{McLeod1996} might in
principal have used curvature of field (rather than astigmatism) to
align the Whipple 1.2-m, but there is a potential degeneracy with a
tilted instrument.  Following \citeauthor{Schroeder1987} and
\citeauthor{McLeod1996}, we find an associated wavefront,

\begin{equation}
\label{eq:cof_s2}
G^{COF} = G^{COF}_{Seidel}  \sigma^2      \rho^2
       + G^{COF}_{decenter} \sigma \ell   \rho^2 \cos(\theta - \phi_{\ell})
       + G^{COF}_{tilt}     \sigma \alpha \rho^2 \cos(\theta - \phi_{\alpha}).
\end{equation}

The first term on the right hand side gives the symmetric Seidel
curvature of field typical of an aligned two mirror telescope, and the
next two give the defocus patterns typical of a decenter or tilt of
the secondary.  The coefficients $G^{COF}_{Seidel}$,
$G^{COF}_{decenter}$, and $G^{COF}_{tilt}$ again depend upon the radii
of curvature, $R_i$ and conic constants, $K_i$ of the two mirrors, the
indices of refraction of the air preceding the mirrors, $n_i$, the
positions, $s_i$ and magnifications, $m_i$ of the object for each
mirror, and the distance from the primary mirror to
the secondary, $W$.  The details of the derivation of the above are
given in appendix C.

\begin{figure}[htb]
\vspace{3.0 truein}
\includegraphics{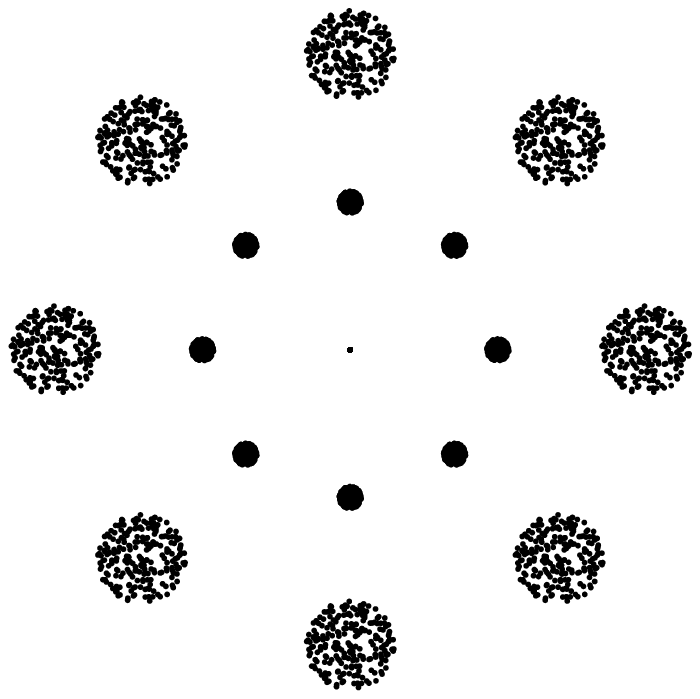}
\includegraphics{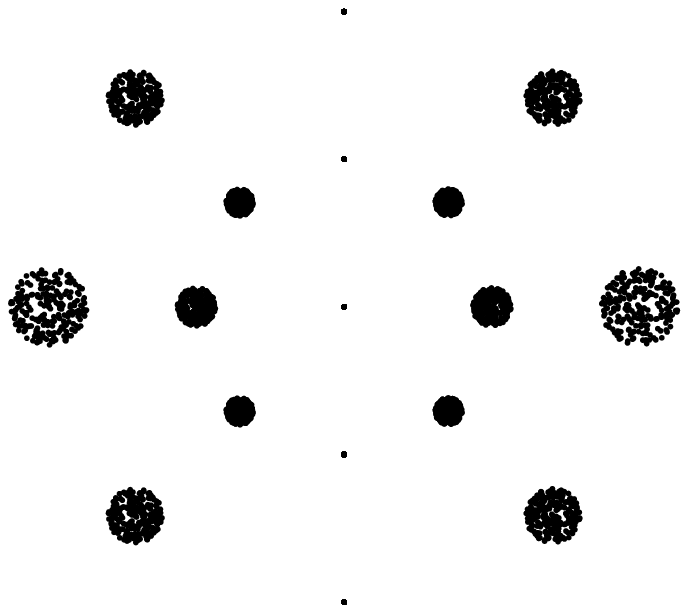}
\caption{Curvature of field patterns.  (a) Seidel COF typical of an
  aligned telescope.  (b) COF field indicating a tilt and (or)
  decenter of the secondary along the x axis. \label{fig:cof_s2}}
\end{figure}

The Seidel term varies as the square of the field radius $\sigma$ and
varies as $\rho^2$ on the pupil.  But the dependence on pupil position
is exactly the same as that of defocus, which is a first-order
aberration.  Anticipating the nomenclature introduced in \S
\ref{sec:generic_fifth-order} below, curvature of field might equally
well be called ``third-order defocus,'' but we bow to convention.
The tilt and decenter terms vary linearly with distance from a line
passing through the center of the field, but have the same functional
dependence on pupil coordinates.  This is precisely what one would
expect for a tilted focal plane.  Figure\ \ref{fig:cof_s2}a shows the
point spread function at various points in the field for the Seidel
term in equation \eqref{eq:cof_s2}.  Figure \ref{fig:cof_s2}b shows
the point spread function pattern introduced by either the tilt or decenter
terms in equation \eqref{eq:cof_s2}.

\subsection{distortion}

While distortion is one of the five Seidel aberrations, for many
purposes it can be neglected, since it does not degrade image quality.
Distortion does, however alter the positions of images in the field,
and of particular interest for the measurements of weak gravitational
lensing, it changes the shapes of extended objects.  And most
importantly for the present discussion it may be of some use in
aligning a telescope.

Following the 
conventions of \citeauthor{Schroeder1987} and \citeauthor{McLeod1996}, we find the associated 
wavefront delay for a misaligned two mirror telescope:

\begin{align}
\label{eq:distortion_s2}
G^{distortion} & = G^{distortion}_{Seidel}   \sigma^3 \rho \cos(\phi - \theta) \\
\nonumber             & + G^{distortion}_{decenter,\sigma} 
              \sigma^2 \ell \cos(\theta - \phi_\ell) \rho \cos(\phi - \theta)
              + G^{distortion}_{decenter,\rho} 
              \sigma^2 \ell \rho \cos(\phi - \phi_{\ell}) \\
\nonumber             & + G^{distortion}_{tilt,\sigma} 
              \sigma^2 \alpha \cos(\theta - \phi_\alpha)\rho \cos(\phi - \theta)
              + G^{distortion}_{tilt,\rho} 
              \sigma^2 \alpha \rho \cos(\phi - \phi_{\alpha})
\end{align}

\noindent
The Seidel term on the right varies as the cube of field angle
$\sigma$ and as $\rho \cos \phi$ and (or) $\rho \sin \phi$ on the
pupil.  Distortion differs from coma, astigmatism and curvature of
field in having two distinct misalignment aberration patterns rather
than only one.  We will encounter several similar pairs of
misalignment aberration patterns when we consider fifth-order
misalignment patterns in \S4. The two terms are distinguished by
whether the direction of the tilt, $\vec \alpha$ or decenter, $\vec
\ell$ enters in a dot product with the field position, $\vec \sigma$
or the pupil position, $\vec \rho$.  We use ``$\sigma$'' and
``$\rho$'' to label the two alternatives.

The $\sigma$  terms have the same functional dependence on
pupil coordinates as the Seidel term, but are proportional to the square
of the field angle $\sigma$ and the cosine of its polar coordinate, $\theta$.
These produce a field distortion pattern directed radially outward,
but with a magnitude that depends on the product of the field angle and
its projection onto the decenter or tilt.

The $\rho$ terms also have the same functional dependence on
pupil coordinates as the first, but vary only as the square field
angle $\sigma$.  The direction of the distortion is that of the decenter or
tilt, but its magnitude increases outward as the square of the
distance from the center of the field.

The three distortion patterns are shown in figure\ \ref{fig:distortion_s2}.
Details of the derivation of the coefficients are given in appendix
D.

Each of the two distinct misalignment distortion patterns is
characterized by a two-vector.  Were one able to measure those vectors
with the same accuracy as the two-vectors that characterize coma and
astigmatism one might in principle use {\it only} distortion measurements 
to align a two mirror telescope.

\begin{figure}[htb]
\vspace{2.5 truein}
\includegraphics{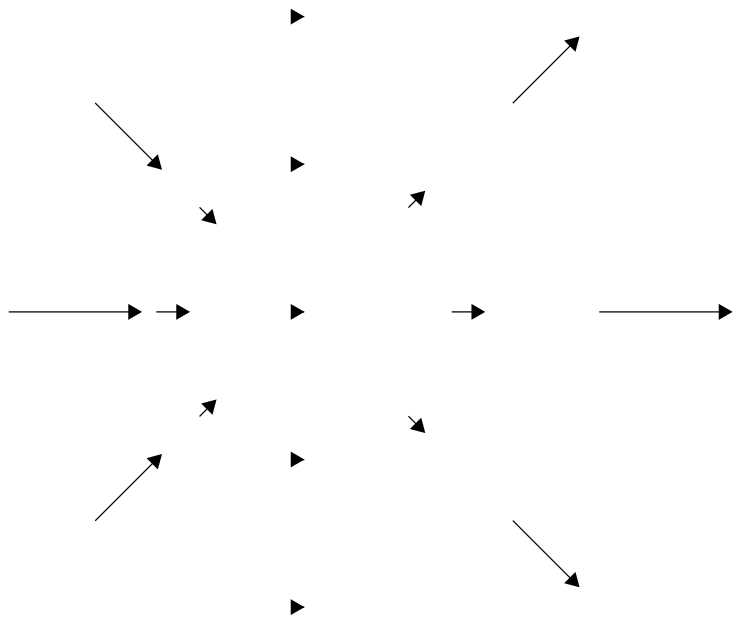}
\includegraphics{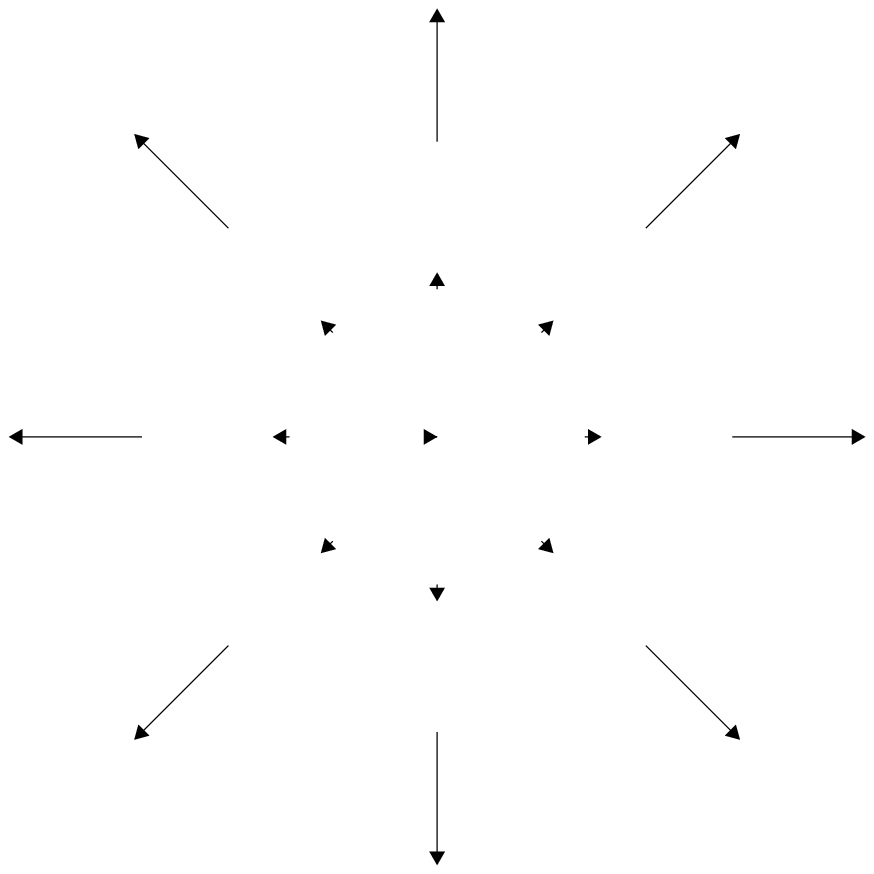}
\includegraphics{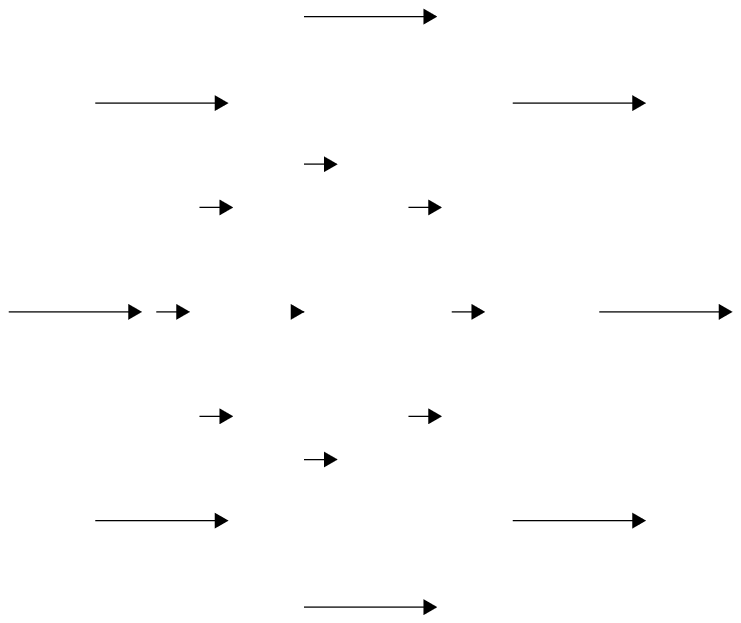}
\caption{Distortion field patterns.  (a) The ``$\sigma$'' distortion
  pattern indicating a tilt and (or) decentering of the secondary
  along the x axis.  
  (b) Seidel distortion typical of an aligned telescope.
  (c) The ``$\rho$'' distortion
  pattern indicating a tilt and (or) decentering of the secondary
  along the x axis.  \label{fig:distortion_s2}}
\end{figure}

\subsection{spherical aberration}
Spherical aberration is the fifth of the Seidel aberrations and
is constant across the field.  
Tilts and decenters do not produce
asymmetric spherical aberration patterns.

\subsection{generalization} 
\label{subsec:generalization}

In the preceding subsections the third-order aberrations were cast so
as to display explicitly the azimuthal dependence upon
pupil position, $\vec \rho$ and field angle, $\vec
\sigma$.  These can be recast more compactly and transparently in
vector form.  For example the Seidel distortion term above varies as
$(\vec\sigma\cdot\vec\sigma)(\vec\sigma\cdot\vec\rho)$.

Suppose a single mirror $i$ serves as its own pupil.  The wavefront
delay $G^{3rd}$ for a ray that intercepts the mirror at position
$\vec\varpi$ and that makes
an angle $\vec\psi$ with the axis of the mirror is given by\footnote{We reserve $\vec{\rho}$ for the position on
  the pupil when the optic is despaced from the pupil.}

\begin{align}
\label{eq:3rd_order_surface_s2}
G^{3rd} &= W_{040}(\frac{\vec{\varpi}}{R} \cdot
\frac{\vec{\varpi}}{R}) (\frac{\vec{\varpi}}{R} \cdot
\frac{\vec{\varpi}}{R}) + W_{131}(\vec{\psi} \cdot
\frac{\vec{\varpi}}{R}) (\frac{\vec{\varpi}}{R} \cdot
\frac{\vec{\varpi}}{R}) \\ \nonumber & + W_{222}(\vec{\psi} \cdot
\frac{\vec{\varpi}}{R})(\vec{\psi} \cdot \frac{\vec{\varpi}}{R}) +
W_{220}(\vec{\psi} \cdot \vec{\psi}) (\frac{\vec{\varpi}}{R} \cdot
\frac{\vec{\varpi}}{R}) \\ \nonumber & + W_{311}(\vec{\psi} \cdot
\vec{\psi})(\vec{\psi} \cdot \frac{\vec{\varpi}}{R})
\end{align}

\noindent 
where $W_{040}$ is the spherical aberration coefficient, $W_{131}$ is
the coma coefficient, $W_{222}$ is the astigmatism coefficient,
$W_{220}$ is the curvature of field coefficient, and $W_{311}$ is the 
distortion coefficient.  These aberration
coefficients depend only on the curvature of the mirror, $R$, the conic
constant of the mirror, $K$, the magnification of the mirror, $m$,
the position of the object for the mirror, $s$, and the index of refraction 
immediately preceding the mirror, $n$\footnote{We adopt the convention of 
\citet{Schroeder1987} whereby rays traveling in opposite directions 	
encounter oppositely signed indices of refraction; a ray incident on the 
primary mirror through air ($n=1$) will encounter a negative index of 
refraction ($n=-1$) once reflected and traveling towards the secondary. 
Therefore, although this discussion deals primarily with mirrors and thin 
lenses, the index of the refraction of the material surrounding the optics 
cannot be entirely ignored.}.  
In \S 5.1 of his book 
\citet{Schroeder1987}
shows that the aberrations which vary linearly with ray height on the optic
must be zero for conic section optics which serve as their own pupils.  
We therefore set $W_{311}$ to zero and ignore it in the following discussion.
The values for the other coefficients are given in Table \ref{tab:Wklm}.

\begin{deluxetable}{@{\extracolsep{\fill}}cccc}
     \tablewidth{0.75\textwidth}
     \tablecaption{Aberration coefficient values for the optic-centered pupil.\label{tab:Wklm}}
     \tablecolumns{4}
     \tablehead{
          \colhead{$W_{040}$} &
          \colhead{$W_{131}$} &
          \colhead{$W_{222}$} &
          \colhead{$W_{220}$} 
     }
     \startdata
	$\frac{nR}{4}((\frac{R}{s}-1)^2 + K)$ &
	$-nR(\frac{R}{s}-1)$ &
	$nR$ &
	$0$       \\ 
     \enddata
\end{deluxetable}

Following \citet{Schroeder1987}, one finds that if the pupil is offset by $W$ along the axis of the mirror, the
position at which a ray lands on the mirror, $\vec\varpi$ depends upon
its position on the pupil, $\vec\rho$ and the angle the
chief ray makes to the pupil normal, $\vec\sigma$ (which is defined to be the field
angle), and upon $W$.  The angle that the ray makes with the axis of
the mirror, $\vec\psi$ depends upon field angle $\vec\sigma$ and the pupil
offset, $W$.

If the mirror is tilted by angle $\vec\alpha$, it changes the angle
$\vec\psi$ that a ray makes with the mirror's axis.  And if the mirror
is decentered from the optical axis (which is perpendicular to and
centered on the pupil) by $\vec\ell$, it changes
both the position at which a ray lands on the
mirror, $\varpi$ and the angle $\vec\psi$ that the ray makes with
respect to the mirror's axis.  

The net effect of a pupil offset or a misalignment of the mirror relative
to the pupil is to shift the position $\vec\varpi$ at which a given
ray strikes the mirror and to change the angle $\vec{\psi}$ that the
ray makes with the axis of the mirror.  The transformations from pupil
coordinates, $\vec\rho$ and field angle, $\vec \sigma$ 
to mirror coordinates and mirror angle for a
mirror despaced by an amount $W$ and decentered and tilted by
$\vec{l}$ and $\vec{\alpha}$ are then\footnote{In deriving equations \eqref{eq:surface2pupil_s2a} and \eqref{eq:surface2pupil_s2b} we assume that all of the effects of the curvature of an optic on aberrations are embodied in equation \eqref{eq:3rd_order_surface_s2}, and we subsequently treat displacements, $W$, decenters, $\vec\ell$ and tilts, $\vec\alpha$ of an optic as displacements, decenters, and tilts of the flat surface defined by the plane containing the vertex of that optic.  This is equivalent to using a first order approximation of pupil and field coordinates.  While this assumption reproduces the stop shift formulae given in \citet{Schroeder1987} and \citet{Wilson1996}, it is not strictly correct in the cases of large or very curved mirrors, large fields, or cases where the stop is quite close to the optic.  In particular, when a telescope's primary mirror is the aperture stop, the computed aberrations are only correct when $W$ for the primary mirror is set equal to zero, even though the physical aperture is defined by the edge of the optic, which may actually be a finite distance from the optic's vertex.  Setting $W$ equal to zero in this special case is natural, as equation \eqref{eq:3rd_order_surface_s2} completely describes the aberrations introduced by an optic that is its own pupil, without any need for modification.  However, this inconsistency would seem to indicate that the assumptions that predicate equations \eqref{eq:surface2pupil_s2a} and \eqref{eq:surface2pupil_s2b} may need to be modified in order to compute higher order aberrations.}

\begin{align}
\label{eq:surface2pupil_s2a}
     \vec{\psi} &=(1-\frac{W}{s})\vec{\sigma} - (\vec{\alpha} + \frac{\vec{l}}{s}) \\
\label{eq:surface2pupil_s2b}
     \vec{\varpi} &= (\vec{\rho} - W \vec{\sigma}) - \vec{l}.
\end{align}

An offset of the pupil from the mirror by an amount $W$ causes what was
Seidel spherical aberration to manifest itself as a combination of
spherical aberration, coma, astigmatism, curvature of field and
distortion, all of which have the symmetric Seidel field dependence.
Likewise what was coma manifests itself as a combination of coma,
astigmatism, curvature of field and distortion.  This cascade downward
from spherical to coma to astigmatism and curvature of field and
finally to distortion is embodied in the ``stop shift'' formulae
(e.g. \citealt{Wilson1996}).

Decenterings and tilts of the mirror relative to the pupil also
produce cascades.  For an axially symmetric telescope we may assume
that these tilts and decenters are small and ignore terms that are
quadratic and higher in either or both.  In the aberration patterns
associated with the surviving linear terms, a field angle vector
$\vec\sigma$ is replaced by either a tilt, $\vec\alpha$ or a decenter,
$\vec \ell / R$.  The field angle exponents for the misalignment
aberration patterns are therefore smaller by one than those of the
corresponding Seidel aberrations.

Table \ref{tab:3rd_order} gives aberration patterns that
arise when a mirror is offset by $W$ with respect to its pupil,  
and decentered and tilted by small amounts.

The different signs for odd and even numbered mirrors arise from the
fact that the chief ray for a given optic may be traveling in the
opposite angular direction than the chief ray on the pupil for the primary
mirror, which here determines the field angle $\vec \sigma$.  Put otherwise, the sign accounts for the fact that preceding mirrors flipped the images.  It does not account for the changing indices of refraction which are contained in the $W_{klm}$ coefficients.  The
  $\sigma$-type distortion patterns have factors
  $\vec{\sigma}\cdot\frac{\vec{l}}{R}$ or
  $\vec{\sigma}\cdot\vec\alpha$.  The $\rho$-type distortion patterns
  have factors $\vec{\rho}\cdot\frac{\vec{l}}{R}$ or
  $\vec{\rho}\cdot\vec\alpha$.

\begin{center}
\begin{deluxetable}{l   |   l    |    l    |   l }
     \tabletypesize{\small} 
     \tablewidth{0pt}
     \tablecaption{Symmetric and asymmetric aberration patterns for mirrors
     offset, decentered and tilted with respect to the pupil. \label{tab:3rd_order}}
     \tablecolumns{4}
     \tablehead{
          \colhead{aberration} &
          \colhead{pupil offset $W$} &
          \colhead{decenter $\vec\ell$} &
          \colhead{tilt $\vec\alpha$}
     }
     \startdata
          spherical 
               & $ (\frac{\vec{\rho}}{R}\cdot\frac{\vec{\rho}}{R})(\frac{\vec{\rho}}{R}\cdot\frac{\vec{\rho}}{R})  \times W_{040}$ 
               & 
               & \\ [1.9ex]
          
          coma  
               & $(\frac{\vec{\rho}}{R}\cdot\frac{\vec{\rho}}{R})(\frac{\vec{\rho}}{R}\cdot\vec{\sigma})  \times \mp \tablenotemark{a} \big[$ 
               & $(\frac{\vec{\rho}}{R}\cdot\frac{\vec{\rho}}{R})(\frac{\vec{\rho}}{R}\cdot\frac{\vec{l}}{R})  \times \big[$ 
               & $(\frac{\vec{\rho}}{R}\cdot\frac{\vec{\rho}}{R})(\frac{\vec{\rho}}{R}\cdot\vec{\alpha})  \times \big[$  
               \\ [0.9ex]
               
               & $\qquad   \left(\frac{W}{s} -1 \right)W_{131}$ 
               & $\qquad  - \left(\frac{R}{s}\right)W_{131} $ 
               & $\qquad - W_{131}$ 
               \\[0.9ex]
               
               & $ \qquad + 4\frac{W}{R} W_{040} \qquad \big]$ 
               & $ \qquad - 4W_{040} \qquad \big]$ 
               & $ \qquad \qquad \qquad \qquad  \big] $ \\[1.9ex]

          astigmatism 
               & $(\frac{\vec{\rho}}{R}\cdot\vec{\sigma})(\frac{\vec{\rho}}{R}\cdot\vec{\sigma})  \times \big[$ 
               & $(\frac{\vec{\rho}}{R}\cdot\vec{\sigma})(\frac{\vec{\rho}}{R}\cdot\frac{\vec{l}}{R})  \times \pm\tablenotemark{a} \big[$
               & $(\frac{\vec{\rho}}{R}\cdot\vec{\sigma})(\frac{\vec{\rho}}{R}\cdot\vec{\alpha})  \times \pm \tablenotemark{a} \big[$  
               \\ [0.9ex]
              
               & $\qquad   \left(\frac{W}{s} - 1 \right)^2W_{222}$ 
               & $\qquad  2\left(\frac{R}{s}\right)\left(\frac{W}{s} - 1 \right)W_{222}$
               & $\qquad  2\left(\frac{W}{s} -1 \right)W_{222}$ 
               \\[0.9ex]
                             
               & $\qquad + 2 \frac{W}{R}\left(\frac{W}{s} - 1\right)W_{131}$ 
               & $\qquad + 2 \left(\frac{2W}{s} - 1 \right) W_{131}$
               & $\qquad + 2 \frac{W}{R} W_{131}$  
               \\ [0.9ex]
               
               & $ \qquad + 4\frac{W^2}{R^2} W_{040} \qquad \big]$
               & $ \qquad + 8 \frac{W}{R}W_{040} \qquad \big]$
               & $ \qquad \qquad \qquad \qquad  \big] $ \\ [1.9ex]
        
          COF    
               & $(\frac{\vec{\rho}}{R}\cdot\frac{\vec{\rho}}{R})(\vec{\sigma}\cdot\vec{\sigma})  \times \big[$
               & $(\frac{\vec{\rho}}{R}\cdot\frac{\vec{\rho}}{R})(\vec{\sigma} \cdot \frac{\vec{l}}{R})  \times \pm \tablenotemark{a} \big[$
               & $(\frac{\vec{\rho}}{R}\cdot\frac{\vec{\rho}}{R})(\vec{\sigma} \cdot \vec{\alpha})  \times \pm \tablenotemark{a} \big[$
               \\ [0.9ex]
               
               & $\qquad  \left(\frac{W}{s} -1 \right)^2W_{220}$
               & $\qquad  2\left(\frac{R}{s}\right)\left(\frac{W}{s} - 1 \right)W_{220}$ 
               & $\qquad  2\left(\frac{W}{s} -1 \right)W_{220}$  
               \\ [0.9ex]

               & $\qquad + \frac{W}{R}\left(\frac{W}{s} -1 \right)W_{131}$
               & $\qquad + \left(\frac{2W}{s} - 1 \right) W_{131}$ 
               & $\qquad + \frac{W}{R} W_{131}$  
               \\ [0.9ex]

               & $ \qquad + 2\frac{W^2}{R^2} W_{040} \qquad \big]$
               & $ \qquad + 4\frac{W}{R}W_{040} \qquad \big]$ 
               & $ \qquad \qquad \qquad \qquad  \big] $ \\[1.9ex]

          distortion  
               & $(\frac{\vec{\rho}}{R}\cdot\vec{\sigma})(\vec{\sigma}\cdot\vec{\sigma})  \times \mp \tablenotemark{a}\big[$ \tablenotemark{b}
               & $(\frac{\vec{\rho}}{R}\cdot\frac{\vec{l}}{R})(\vec{\sigma}\cdot\vec{\sigma})  \times \big[$ 
               & $(\frac{\vec{\rho}}{R}\cdot\vec{\alpha})(\vec{\sigma}\cdot\vec{\sigma})  \times \big[$   
               \\ [0.9ex]

               & $\qquad    2\frac{W}{R}\left(\frac{W}{s} - 1\right)^2 W_{220}$
               & $\qquad    -2\left(\frac{W}{s} - 1\right)^2W_{220}$ 
               &   
               \\ [0.9ex]

               & $\qquad + 2\frac{W}{R}\left(\frac{W}{s} - 1\right)^2 W_{222} $
               & $\qquad - 2\left(\frac{W}{s}\right)\left(\frac{W}{s} - 1\right)W_{222}$ 
               & $\qquad    -2\frac{W}{R}\left(\frac{W}{s} -1 \right)W_{222} $  
               \\ [0.9ex]
                             
               & $ \qquad +3\frac{W^2}{R^2}\left(\frac{W}{s} - 1 \right)W_{131} $
               & $ \qquad - 2\frac{W}{R}\left(\frac{3W}{2s} - 1\right)W_{131} $ 
               & $ \qquad - \frac{W^2}{R^2}W_{131} $
               \\ [0.9ex]

               & $ \qquad +4 \frac{W^3}{R^3} W_{040} \qquad \big]$
               & $ \qquad - 4\frac{W^2}{R^2}W_{040}  \qquad \big]$ 
               & $ \qquad \qquad \qquad \qquad  \big] $ \\[1.9ex]
                          
               & 
               & $(\frac{\vec{\rho}}{R}\cdot\vec{\sigma})(\vec{\sigma}\cdot\frac{\vec{l}}{R})  \times \big[$ 
               & $(\frac{\vec{\rho}}{R}\cdot\vec{\sigma})(\vec{\sigma}\cdot\vec{\alpha})  \times \big[$  
               \\ [0.9ex]
             
               & 
               & $\qquad    -4\left(\frac{W}{s}\right)\left(\frac{W}{s} - 1\right)W_{220} $ 
               & $\qquad    -4\frac{W}{R}\left(\frac{W}{s} -1 \right)W_{220} $  
               \\ [0.9ex]
             
               & 
               & $ \qquad - 2\left(\frac{W}{s} -1 \right) \left(\frac{2W}{s} - 1\right)W_{222}$ 
               & $\qquad -2\frac{W}{R}\left(\frac{W}{s} -1 \right)W_{222} $ 
               \\ [0.9ex]

               & 
               & $ \qquad - 4\frac{W}{R}\left(\frac{3W}{2s} - 1\right)W_{131} $ 
               & $ \qquad - 2\frac{W^2}{R^2}W_{131}$
               \\ [0.9ex]

               & 
               & $ \qquad - 8\frac{W^2}{R^2}W_{040} \qquad \big]$ 
               & $ \qquad \qquad \qquad \qquad  \big] $ \\             
\enddata
\tablenotetext{a}{\footnotesize The upper sign is for a primary, tertiary, or other
  odd numbered mirror.  The lower sign is for a secondary or even
  numbered mirror.}  
\tablenotetext{b}{\footnotesize The $W_{311}$ term has been omitted as it is equal to zero.}  
\end{deluxetable}
\end{center}

\subsection{application to 2-mirror telescopes and 2.5-mirror telescopes}
\label{sec:aligning_2.5-mirror}

Table \ref{tab:3rd_order} of the previous subsection gives the third-order
aberrations for a single mirror with a pupil offset by $W$ along the
optical axis.  The aberrations for a 2-mirror telescope are found by
computing the elements of two such tables, one for the primary and one
for the secondary, and adding.

Before one can use Table \ref{tab:3rd_order} to compute aberrations, one 
must first know the location of the center of the field as it is defined by the 
pointing of the pupil.  If the center of pointing is unknown then one must add 
an additional vector variable to the table, $\vec{m}$, which maps the true 
field coordinate, $\vec{\sigma}$  to that adopted for measurement, 
$\vec{\sigma^{\prime}}$ via the relation 
$\vec{\sigma} \rightarrow\vec{\sigma^{\prime}} - \vec{m}$.  
This focal plane decenter will mathematically 
map Seidel terms to misalignment terms, but it will not physically aberrate the 
images. For the sake of simplicity, we here assume that the center of pointing 
is known and treat a decentered focal plane as a `complication' in \S
\ref{sec:complications}.

For many two-mirror telescopes the pupil is coincident with the
primary.  If one also takes the primary to define the pointing of the
telescope, one can take the pupil offset, $W_1$, the mirror decenter,
$\vec\ell_1$ and the mirror tilt, $\vec\alpha_1$ all to be zero.  The
misalignment patterns are then due entirely to the decenter,
$\vec\ell_2$ and tilt, $\vec\alpha_2$ of the secondary.  One need only
measure two of the five misalignment patterns in Table
\ref{tab:3rd_order} to align the telescope.  If the aperture stop (and 
consequently the pointing) instead
coincide with the secondary mirror as in \citet{NoetheGuisard2000}, the
misalignments of the secondary mirror may be treated as identically zero
and only the misalignments of the primary mirror relative to its pupil need 
to be considered.  In this case as well, one need only
measure two of the five misalignment patterns in Table
\ref{tab:3rd_order} to align the telescope.  There is a third case where 
the aperture stop is independent from either of the mirrors.  In this case
we would treat the aperture stop as a third optical element and analyze
this telescope configuration as a three mirror telescope.

\citet{McLeod1996} measures coma and astigmatism in his two mirror 
telescope.  He might in principle have used the two
distortion patterns, but these would require a high precision
astrometric catalog (perhaps using galaxy positions to avoid the
effects of stellar proper motions).  Moreover as the misalignment
distortion patterns vary as a higher powers of field angle than the
misalignment astigmatism and coma patterns, they might be expected to
have smaller amplitudes.  McLeod might also have used curvature of
field, although here there is the danger that the detector might be
tilted with respect to the primary.  Or, had he been feeling
particularly masochistic, he might have measured all five patterns,
for the sake of redundancy.

With their folding flats, the Magellan telescopes in their Nasmyth
configuration qualify as 2.5-mirror telescopes.  One need not worry
about the decentering of the tertiary but one must measure and correct
for its tilt, $\vec \alpha_3$.  The alignment procedure described
by \citet{PalunasFloyd2010} adds the curvature of field misalignment
pattern (equivalent to a focal plane tilt) to those of coma and
astigmatism.

\section{Aligning 3-mirror telescopes using distortion patterns}
\label{sec:aligning_3-mirror}

Calculating the aberration patterns for a 3-mirror telescope (say a
three mirror anastigmat, henceforth a TMA) is not quite twice as
difficult as for a 2-mirror telescope.  One applies Table
\ref{tab:3rd_order} to the tertiary and finds misalignment patterns
that depend upon the decenter $\vec\ell_3$ and tilt $\vec\alpha_3$ of
the tertiary.  If the stop is coincident with the primary, only the
secondary and tertiary contribute to the five third-order
misalignment patterns in Table \ref{tab:3rd_order}.  The patterns are linear in 
the decenter and tilt vectors, so that the 
combined wavefront gives the same five patterns, each characterized by a 
new pattern 2-vector.  Each pattern 2-vector, $\vec\mu$, is a linear 
combination of the four misalignment 2-vectors, the tilts $\vec \alpha_2$ 
and $\vec \alpha_3$, and the decenterings $\vec \ell_2$ and $\vec \ell_3$.  
Likewise if the stop is coincident with the secondary or tertiary mirrors, only 
tilts and decenters of the other two mirrors contribute.

If coma, astigmatism and curvature of field are the only aberrations
that adversely affect science, there is a two-dimensional ``subspace
of benign misalignment'' for which the coma, astigmatism and curvature
of field misalignment patterns are all zero in the limit of small
misalignments.  But if one does not bring additional information to
bear, one runs the risk of drifting increasingly far from perfect
alignment.  This would produce large misalignment distortion, and
ultimately, fifth-order misalignment patterns and third-order patterns
that depend quadratically on the mirror tilts and decenters.

To measure the misalignment distortion patterns one would need either 
overlapping fields of images \citep{Sudol2011} or pre-existing astrometry. If 
one uses overlapping fields, one risks changing the misalignment between 
pointings, thus rendering distortion of limited use for measuring alignment at 
a given pointing.  If one relies on pre-existing astrometry, the accuracy with 
which the distortion patterns could be measured would then be limited by 
the accuracy of the astrometric catalog. Moreover, regardless the technique 
used to measure the aberrations, the misalignment distortion patterns might 
be expected to have smaller amplitudes than misalignment astigmatism and 
coma as the misalignment distortion patterns vary as a higher power of field 
angle. This raises the question of whether one might use fifth-order 
aberration patterns to keep the telescope aligned.

\section{Generic fifth-order aberration patterns}
\label{sec:generic_fifth-order}

\subsection{fifth-order aberrations for a single mirror}

In \S \ref{subsec:generalization} we considered the third-order aberrations produced by a
single mirror with a stop at the mirror.  The generalization to
fifth-order is straightforward.  The fifth-order wavefront delay, $G^{5th}$
for a ray that hits the mirror at position $\vec\varpi$ and that makes
an angle $\vec\psi$ with the axis of the mirror is given by

\begin{align}
\label{eq:5th_order_surface_s4}
G^{5th} & = W_{060} (\frac{\vec\varpi}{R}\cdot\frac{\vec\varpi}{R})
                   (\frac{\vec\varpi}{R}\cdot\frac{\vec\varpi}{R})
                   (\frac{\vec\varpi}{R}\cdot\frac{\vec\varpi}{R}) 
         + W_{151} (\frac{\vec\varpi}{R}\cdot\frac{\vec\varpi}{R}) 
                   (\frac{\vec\varpi}{R}\cdot\frac{\vec\varpi}{R})
                   (\frac{\vec\varpi}{R}\cdot\vec\psi)  \\
\nonumber       & + W_{242} (\frac{\vec\varpi}{R}\cdot\frac{\vec\varpi}{R}) 
                  (\frac{\vec\varpi}{R}\cdot\vec\psi)
                  (\frac{\vec\varpi}{R}\cdot\vec\psi) 
         + W_{240} (\frac{\vec\varpi}{R}\cdot\frac{\vec\varpi}{R}) 
                  (\frac{\vec{\varpi}}{R}\cdot\frac{\vec\varpi}{R}) 
                  (\vec\psi \cdot\vec\psi) \\
\nonumber       & + W_{333} (\frac{\vec{\varpi}}{R}\cdot\vec\psi)
                  (\frac{\vec{\varpi}}{R}\cdot\vec\psi) 
                  (\frac{\vec{\varpi}}{R}\cdot\vec\psi) 
         + W_{331} (\frac{\vec{\varpi}}{R}\cdot\frac{\vec\varpi}{R})
                  (\frac{\vec{\varpi}}{R}\cdot\vec\psi)
                  (\vec\psi \cdot\vec\psi) \\
\nonumber       & + W_{422} (\frac{\vec{\varpi}}{R}\cdot\vec\psi)
                   (\frac{\vec{\varpi}}{R}\cdot\vec\psi)
           (\vec\psi \cdot\vec\psi)  
         + W_{420} (\frac{\vec{\varpi}}{R}\cdot\frac{\vec{\varpi}}{R})
           (\vec\psi \cdot\vec\psi)   
           (\vec\psi \cdot\vec\psi)   \\ 
\nonumber       & + W_{511} (\frac{\vec{\varpi}}{R}\cdot\vec\psi)(\vec\psi \cdot\vec\psi)
           (\vec\psi \cdot\vec\psi) 
\end{align}

Since the pupil is coincident with the mirror, we might equally well
have written the same equations but with position on the mirror
$\vec\varpi$ replaced by position on the pupil $\vec\rho$ and the
angle that a ray makes with the axis of the mirror, $\vec\psi$
replaced by the field angle $\vec\sigma$.  For the remainder of this
subsection we shall take $\vec\rho = \vec\varpi$ and $\vec\sigma = \vec\psi.$

\subsubsection{$W_{511}$: fifth-order distortion}

The $W_{511}$ term has the same variation on the pupil as distortion,
but varies as field angle to the fifth.  We shall refer to this as
``fifth-order distortion.''\footnote{Our nomenclature is driven
  primarily by the functional form of the aberration on the pupil.
  Thus an aberration that varies as $\rho^3\cos\phi$ is referred to as
  coma.  In this scheme third-order (Seidel) coma varies linearly with
  field angle and fifth-order coma varies as the cube of field angle.
  By contrast, the term that varies as $\rho^5\cos\phi$ on the pupil
  is referred to here as second coma or coma-II.  \citet{Hopkins1950} calls
  this term fifth-order coma, but from our pupil oriented perspective
  this term, albeit one of fifth-order, cannot be called coma, which
  can only vary as $\rho^3\cos\phi$.  Curiously, we agree with Hopkins
  in calling the term that varies as $\rho^2\cos^2\phi$ fifth-order
  astigmatism and in calling the term that varies as $\rho\cos\phi$
  fifth-order distortion.  We reserve the term ``Zernike'' aberrations
  for the orthogonalized linear combinations of aberrations described
  here.} 

\subsubsection{$W_{420}$: fifth-order defocus}
The $W_{420}$ term has the same variation on the pupil as defocus, and
produces a point spread function indistinguishable from that of Seidel
curvature of field .  But as this term varies as the fourth power of
field angle rather than quadratically, we call this term ``fifth-order
defocus.''

\subsubsection{$W_{422}$: fifth-order astigmatism}
The $W_{422}$ term has the same variation on the pupil as third-order
astigmatism, and produces a point spread function indistinguishable
from it.  But as this term varies as the fourth power of field angle
rather than quadratically, we call this term ``fifth-order
astigmatism.''  Again as with third-order astigmatism, this term
varies on the pupil as $\rho^2 \cos^2 \phi$.  It can be decomposed
into a term that varies as Zernike astigmatism, $\rho^2 \cos 2\phi$,
and a second that varies as defocus, $\rho^2 \cos^0 \phi$,  

\subsubsection{$W_{331}$: fifth-order coma}

The $W_{331}$ term has the same variation on the pupil as coma, but
varies as field angle cubed.  We call this ``fifth-order coma.''

\subsubsection{$W_{333}$: trefoil}

\begin{figure}[htbp]
\vspace{2.4 truein}
\includegraphics{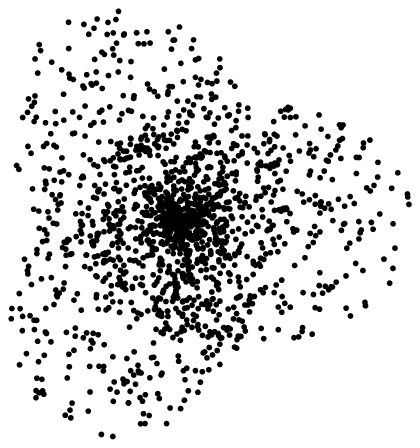}
\includegraphics{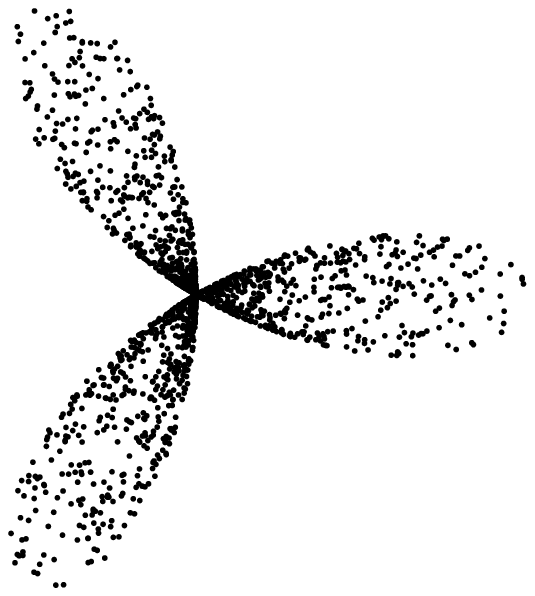}
\includegraphics{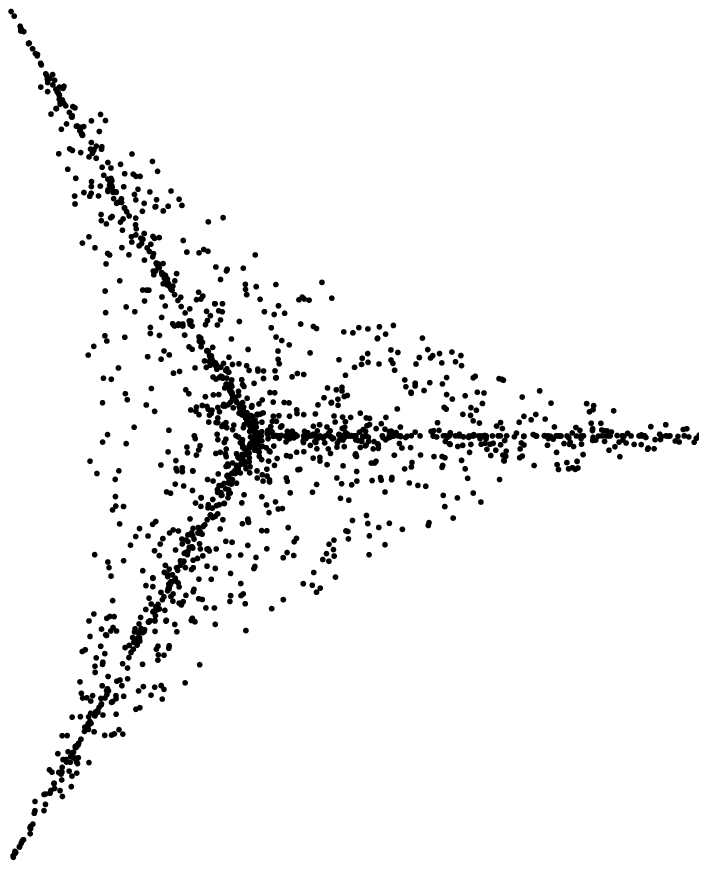}
\caption{Point spread functions due to trefoil
with $G^{tref} = \rho^3 \cos 3\phi$.
Increasing amounts of a
symmetric wavefront $G^{sym} = \rho^3$ have been added
to bring out the three-fold symmetry of the wavefront.
(a) $+\frac{2}{5}G^{sym}$
(b) $+G^{sym}$
(c) $+2G^{sym}$
\label{fig:trefoil_psf_s4}}
\end{figure}

The $W_{333}$ term varies as $\rho^3 \cos^3\phi$ on the pupil and as
field angle cubed.  None of the third-order aberrations has this
behavior on the pupil; here we call it trefoil.  This term can be
decomposed into a term that varies as Zernike trefoil, $\rho^3 \cos
3\phi$, and a second that varies as coma, $\rho^3 \cos \phi$.  Figure \ref{fig:trefoil_psf_s4}
shows the point spread function due to Zernike trefoil with varying
amounts of an aberration that varies as $\rho^3$ added to it.

\subsubsection{$W_{240}$: fifth-order spherical}

The $W_{240}$
has the same variation on the pupil as spherical aberration, but
varies as field angle squared.  We shall refer to this as this
``fifth-order spherical.''  

\subsubsection{$W_{242}$: second astigmatism }

\begin{figure}[htbp]
\vspace{2.0 truein}
\includegraphics{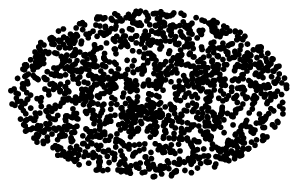}
\includegraphics{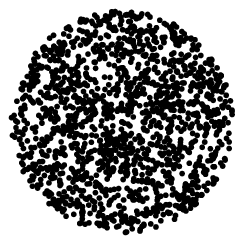}
\includegraphics{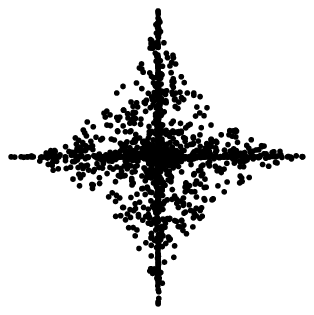}
\caption{Point spread functions for ordinary astigmatism, $G^{astig-I} = \rho^2 cos 2\phi$
and second astigmatism, $G^{astig-II} = \rho^4 cos 2\phi$.  
(a) Ordinary astigmatism $+\frac{1}{4} G^{defocus}$.
(b) Pure ordinary astigmatism.
(c) Pure second astigmatism.
\label{fig:astig2_psf_s4}}
\end{figure}

The $W_{242}$ term varies as $\rho^4 \cos^2\phi$ on the pupil.  The
angular dependence is that of astigmatism but the radial dependence is
quartic not quadratic.  We shall call this ``second astigmatism'' or
``astigmatism-II.''  Figure \ref{fig:astig2_psf_s4} shows the PSF for
second astigmatism next to those for ordinary astigmatism.

\subsubsection{$W_{151}$: second coma }

\begin{figure}[htbp]
\vspace{3.0 truein}
\includegraphics{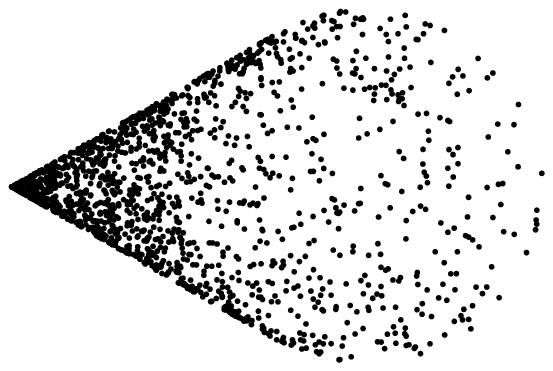}
\includegraphics{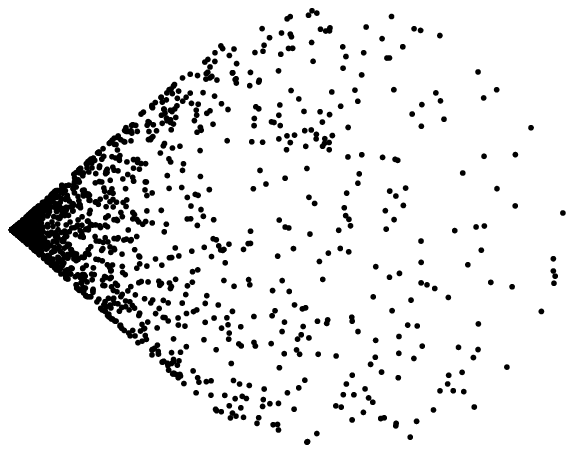}
\caption{(a) Off axis (ordinary) coma PSF for the third-order and
  fifth-order coma patterns. The sine of the apex half angle is $1/2$,
  giving an apex angle of $60^\circ$.
  (b) Off axis second coma PSF.  The sine of the apex half angle
  is $2/3$, giving an apex angle of $83.6^\circ$. \label{fig:coma2_psf_s4}}
\end{figure}

The $W_{151}$ term varies as $\rho^5 \cos\phi$ on the pupil.  The
azimuthal dependence is that of coma but the radial dependence is
quintic not cubic.  We shall refer to this as ``second coma''
or ``coma-II.''    Figure \ref{fig:coma2_psf_s4} shows a PSF produced by second coma
alongside one produced by ordinary (first) coma.  

\subsubsection{$W_{060}$: second spherical }

We shall call the $W_{060}$ term ``second spherical'' or
``spherical-II.''  Were we strictly consistent we would have called
spherical aberration ``second defocus'' and would have called this
``third defocus'' but as Emerson says, a foolish consistency is the
hobgoblin of a small mind.

\subsection{fifth-order aberrations for misaligned systems}

Tilts and decenterings of mirrors produce misalignment patterns
in a manner entirely analogous to those of third-order aberrations.
The dependence upon field and pupil position and the first
order dependence on misalignments are given in Table \ref{tab:5th_order}.  
Equations \eqref{eq:surface2pupil_s2a} and \eqref{eq:surface2pupil_s2b} are used to transform from the positions and angles with respect to the optic, $\vec\varpi$ and $\vec\psi$  of equation \eqref{eq:5th_order_surface_s4} to positions  and angles with respect the pupil, $\vec\rho$ and $\vec\sigma$.\footnote{Third-order aberrations produced by an optic (equation \eqref{eq:3rd_order_surface_s2}) may also contribute to the fifth-order aberrations on the pupil when the pupil and the optic do not coincide.  As noted in \S\S \ref{subsec:generalization} equations \eqref{eq:surface2pupil_s2a} and \eqref{eq:surface2pupil_s2b} are truncated at first order in pupil and field coordinates.  Analogous equations for \eqref{eq:surface2pupil_s2a} and \eqref{eq:surface2pupil_s2b} that have been expanded to third order can map third-order aberrations on the optic to fifth-order aberrations on the pupil.  Other such approximations may also need to be relaxed in order to compute the magnitudes of the fifth-order aberrations.}   Terms with the same dependence on both of the latter are then added.  The  vectors $\vec\mu$ indicate linear combinations of misalignment angle $\vec\alpha$ and decenter $\vec\ell$.  

\begin{deluxetable}{l l l }
     \tablecaption{Fifth-order aberration patterns \label{tab:5th_order}}
     \tablewidth{0pt}
     \tablecolumns{3}
     \tablehead{
          \colhead{aberration} &
          \colhead{symmetric} &
          \colhead{misalignment} 
      }
     \startdata
2nd spherical  
          & $(\vec\rho\cdot\vec\rho)(\vec\rho\cdot\vec \rho)(\vec\rho\cdot\vec\rho)$
          & \\
          
     2nd coma  
          & $(\vec\rho\cdot\vec\rho)(\vec\rho\cdot\vec\rho)(\vec\rho\cdot\vec \sigma)$ 
          & $(\vec\rho\cdot\vec\rho)(\vec\rho\cdot\vec\rho)(\vec\rho\cdot\vec\mu_{cII})$ \\
          
     2nd astigmatism  
          & $(\vec\rho\cdot\vec\rho)(\vec\rho\cdot\vec \sigma)(\vec\rho\cdot\vec \sigma)$ 
          & $(\vec\rho\cdot\vec\rho)(\vec\rho\cdot\vec \sigma)(\vec\rho\cdot\vec \mu_{aII})$\\

     fifth-order spherical  
          & $(\vec\rho\cdot\vec\rho)(\vec\rho\cdot \vec\rho)(\vec \sigma \cdot\vec \sigma)$ 
          & $(\vec\rho\cdot\vec\rho)(\vec\rho\cdot\vec\rho )(\vec \sigma\cdot\vec\mu_{5s})$\\

     trefoil  
          & $(\vec\rho\cdot\vec \sigma)(\vec\rho\cdot\vec \sigma)(\vec\rho\cdot\vec \sigma)$ 
          & $(\vec\rho\cdot\vec \sigma)(\vec\rho\cdot\vec \sigma)(\vec\rho\cdot\vec \mu_{t})$\\

     fifth-order coma  
          & $(\vec\rho\cdot\vec \rho)(\vec\rho\cdot\vec \sigma)(\vec \sigma \cdot\vec \sigma)$ 
          & $(\vec\rho\cdot\vec \rho)(\vec\rho\cdot\vec \sigma)(\vec \sigma \cdot\vec\mu_{5c\sigma})$ \\
          
          &
          & $(\vec\rho\cdot\vec \rho)(\vec \sigma \cdot\vec \sigma)(\vec\rho\cdot\vec\mu_{5c\rho})$\\

     fifth-order astigmatism  
          & $(\vec\rho\cdot\vec \sigma)(\vec\rho\cdot\vec \sigma)(\vec \sigma \cdot \vec \sigma)$ 
          & $(\vec\rho\cdot\vec \sigma)(\vec\rho\cdot\vec \sigma)(\vec \sigma \cdot \vec \mu_{5a\sigma}) $\\ 

          &
          & $(\vec\rho\cdot\vec \sigma)(\vec \sigma \cdot \vec \sigma)(\vec\rho\cdot\vec\mu_{5a\rho}) $\\ 

     fifth-order defocus  
          & $(\vec\rho\cdot\vec\rho)(\vec \sigma \cdot\vec \sigma)(\vec \sigma \cdot\vec \sigma)$    
          & $(\vec\rho\cdot\vec\rho)(\vec \sigma \cdot\vec \sigma)(\vec \sigma \cdot\vec \mu_{5f})$ \\

     fifth-order distortion  
          & $(\vec\rho\cdot\vec \sigma)(\vec \sigma\cdot\vec \sigma)  (\vec \sigma\cdot\vec \sigma)$  
          & $(\vec\rho\cdot\vec \sigma)(\vec \sigma\cdot\vec \sigma)  (\vec \sigma\cdot\vec \mu_{5d\sigma})$ \\

          &
          & $(\vec \sigma\cdot\vec \sigma)(\vec \sigma\cdot\vec \sigma)(\vec\rho\cdot\vec\mu_{5d\rho})$\\
     \enddata
\end{deluxetable}

In Table \ref{tab:5th_order} we give only the dependence upon pupil
position and field angle for the symmetric fifth-order aberration
patterns and their associated misalignment patterns.  For the
third-order aberrations, there was one misalignment pattern each
associated with coma, astigmatism and curvature of field, with two
misalignment patterns associated with distortion.  There are again
distinct $\sigma$ and $\rho$ misalignment patterns for fifth-order
distortion, and now also for fifth-order astigmatism and fifth-order
coma.

\subsection{fifth-order aberration patterns and alignment}

The amplitudes of aberration patterns implicit in Table \ref{tab:5th_order}
provide, at least in principle, additional information for use in
aligning telescopes.  But in all cases there are similarities
between the fifth-order misalignment aberration patterns
and their third-order counterparts.  
The ability to distinguish between the two 
depends, for the first group described below, upon the
sampling of the wavefront, and, for the second
group, upon the sampling of the field. 

\subsubsection{second coma: misalignment}

\begin{figure}[htbp]
\vspace{3.0 truein}
\includegraphics{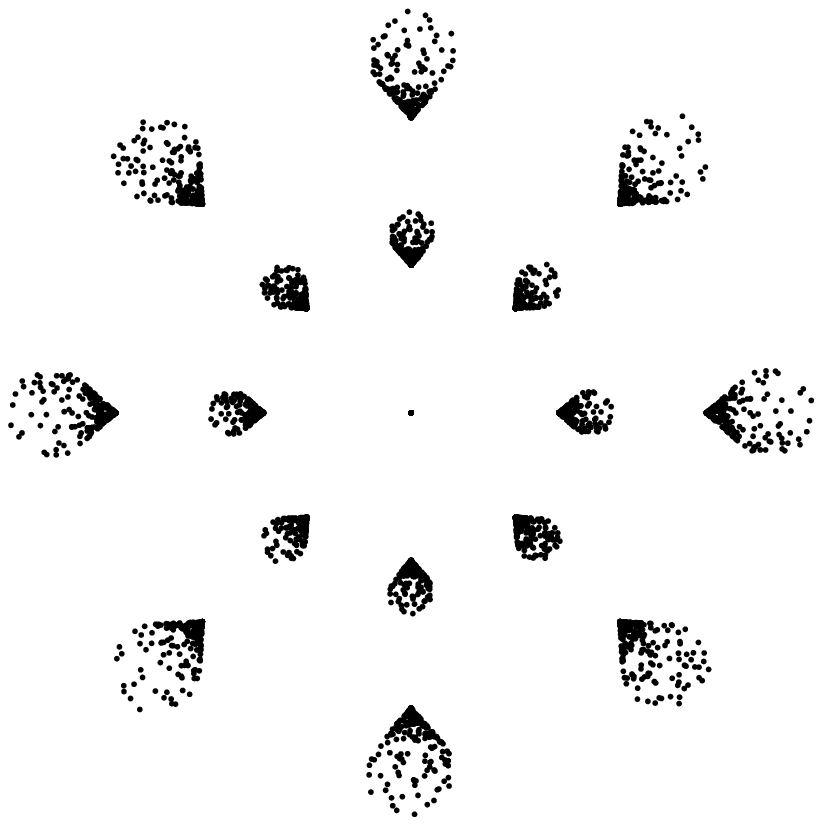}
\includegraphics{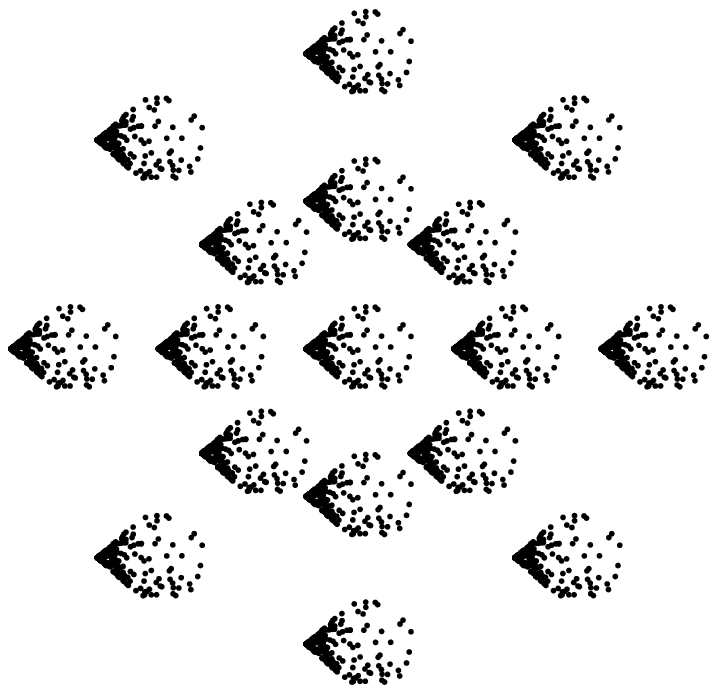}
\caption{Second coma field patterns.  (a) Second coma pattern typical
  of an aligned telescope.  (b) Second coma pattern indicating
  a tilt and (or) decenter of a
  mirror along the x axis. \label{fig:coma2_fp_s4}}
\end{figure}

The misalignment aberration pattern for second coma, shown in Figure
\ref{fig:coma2_fp_s4} is identical to that for ordinary (first) coma
(see Figure \ref{fig:coma_s2}).  The point spread functions for second
coma and ordinary (first) coma have the same azimuthal dependence on
pupil position but different radial dependence.  The ability to
distinguish between the coma-II misalignment pattern and the coma-I
misalignment pattern therefore depends critically upon the sampling of
the wavefront.

\subsubsection{second astigmatism: misalignment}

\begin{figure}[htbp]
\vspace{3.0 truein} 
\includegraphics{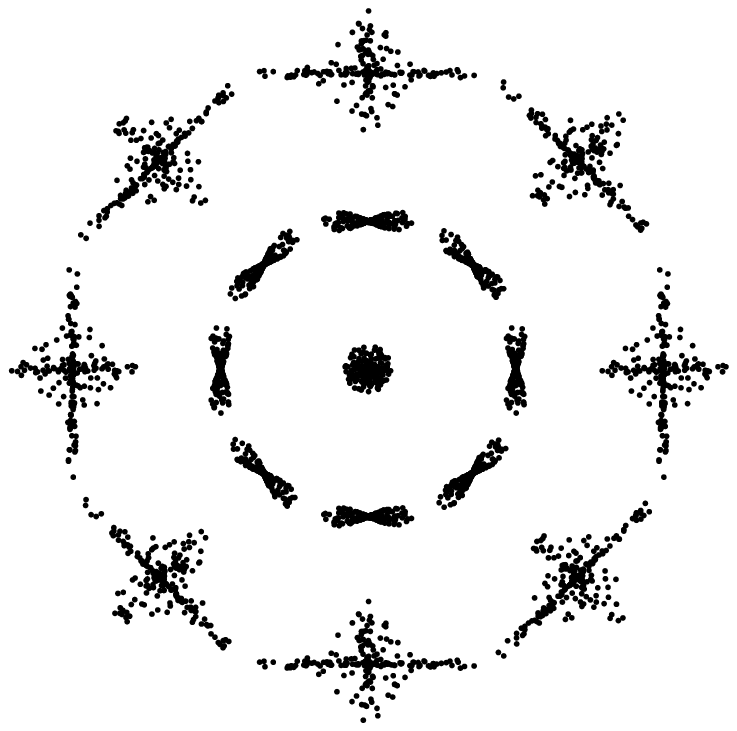}
\includegraphics{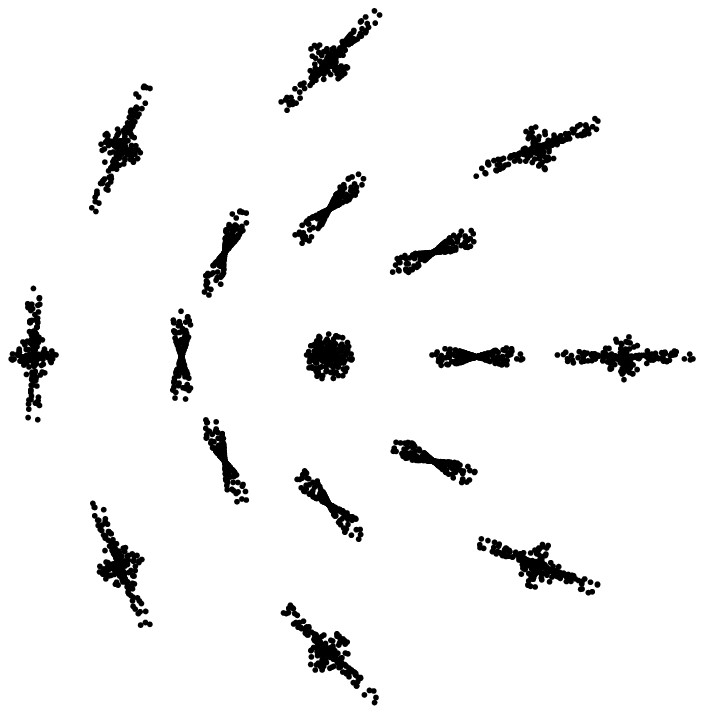} 
\caption{Second astigmatism field patterns.  (a) Second astigmatism
  pattern typical of an aligned telescope.  (b) Second astigmatism
  pattern indicating a tilt and (or)
  decentering of a mirror along the x axis.
  Spherical aberration has been added to bring out the asymmetry
  of the pattern.
  \label{fig:astig2_fp_s4}}
\end{figure}

As with second coma, the misalignment aberration pattern for second astigmatism,
shown in Figure \ref{fig:astig2_fp_s4} is identical to that for
ordinary (first) astigmatism (see Figure \ref{fig:astig_s2}).  The
point spread functions for second astigmatism and ordinary (first)
astigmatism have the same angular dependence on pupil position but
different radial dependence.  The ability to distinguish between the
astigmatism-II misalignment pattern and the astigmatism-I misalignment
pattern again depends critically upon the sampling of the wavefront.

\subsubsection{fifth-order spherical: misalignment}

\begin{figure}[htbp]
\vspace{3.25 truein} \includegraphics{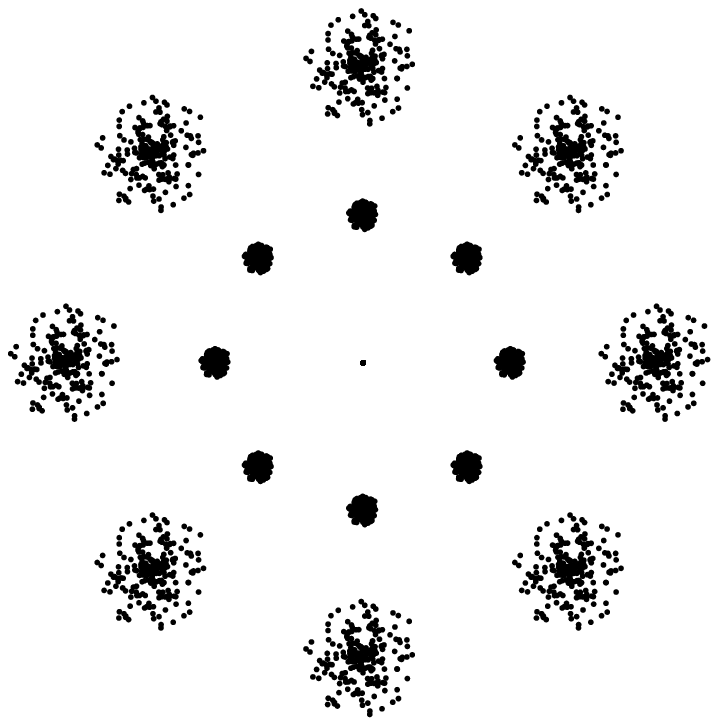}
\includegraphics{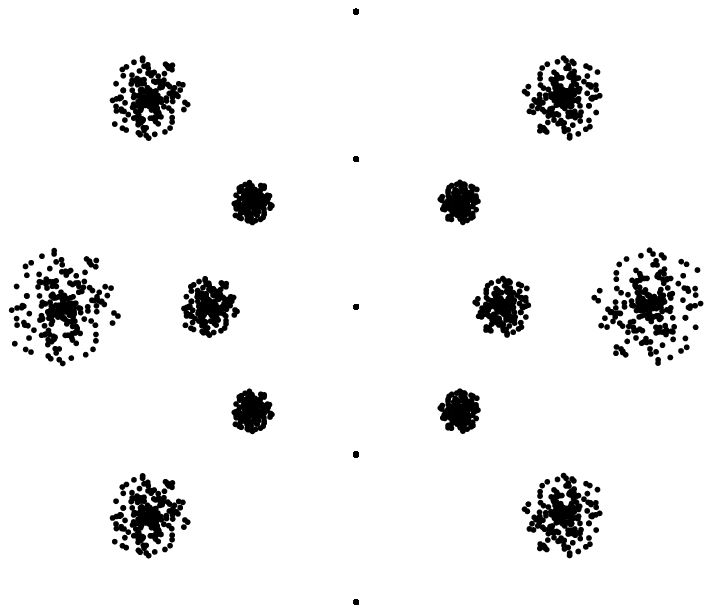}
\caption{Fifth-order spherical field patterns.  (a) Spherical
  aberration pattern typical of an aligned telescope.   
  (b) Spherical aberration indicating
  a tilt and (or) decenter of a mirror
  along the x axis. \label{fig:5th_sph_fp_s4}}
\end{figure}

The aberration patterns for fifth-order spherical, shown in Figure
\ref{fig:5th_sph_fp_s4} are identical to those for those of curvature
of field.  However the point spread functions are identical to spherical 
aberration, which itself has the same angular dependence on the pupil as 
defocus and COF, but a different radial dependence. The ability to 
distinguish between the fifth-order spherical and curvature of field 
misalignment pattern once again depends critically upon the sampling of the 
wavefront.  The ability to distinguish between fifth-order spherical and 
ordinary spherical depends on the sampling of the field.

\subsubsection{trefoil: misalignment}

\begin{figure}[htbp]
\vspace{3.25 truein} 
\includegraphics{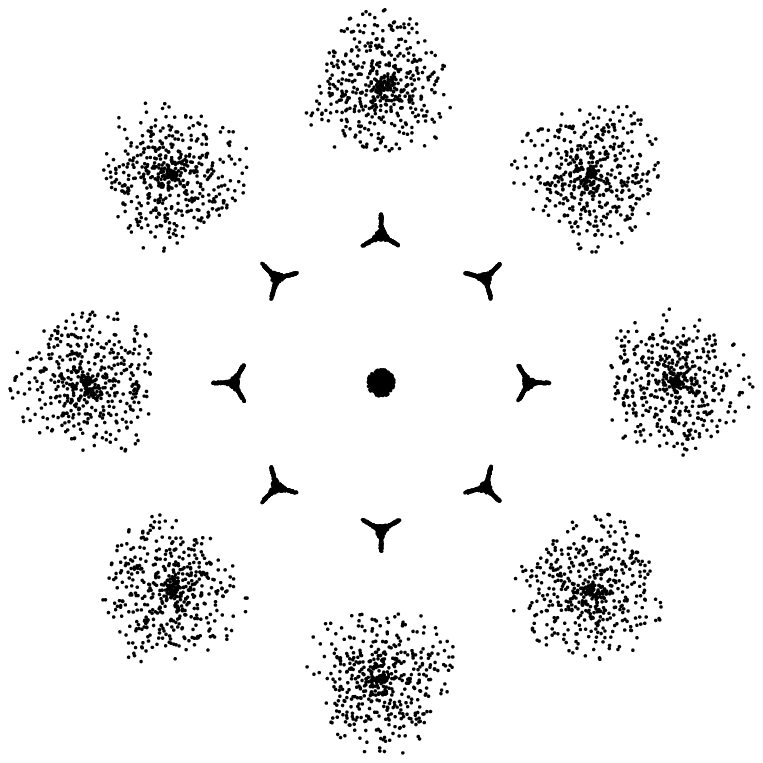}
\includegraphics{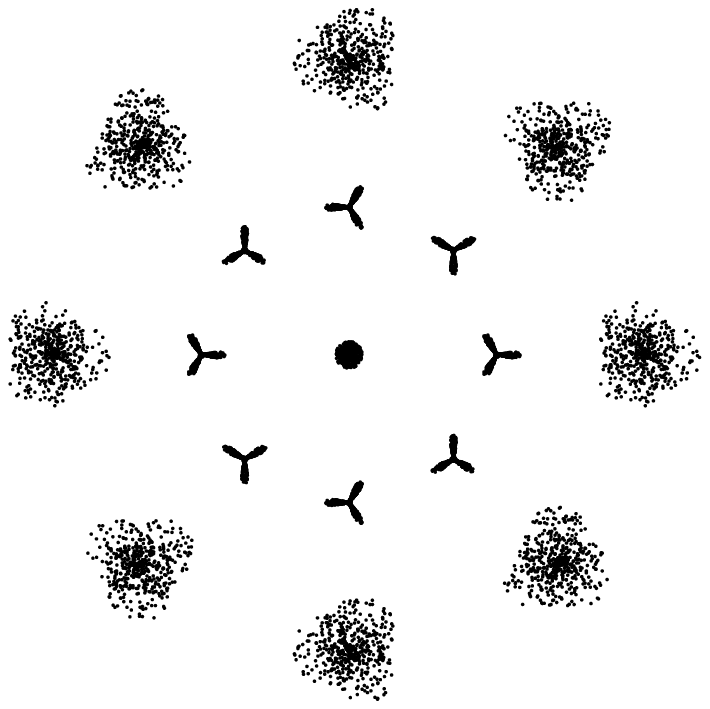} 
\caption{Trefoil field patterns.  (a) Trefoil pattern characteristic
  of an aligned telescope.  (b) Trefoil indicating a tilt and (or)
  decenter of a mirror along the x axis. In both cases a constant
  wavefront varying as $\rho^3$ has been added to bring out the
  orientation of the aberration.\label{fig:trefoil_fp_s4}}
\end{figure}

The PSF for trefoil is quite different from that of any of the Seidel
third-order aberrations, as is its misalignment pattern, shown in
Figure \ref{fig:trefoil_fp_s4}.  

\subsubsection{fifth-order astigmatism: misalignment}

\begin{figure}[htbp]
\vspace{2.0 truein} 
\includegraphics{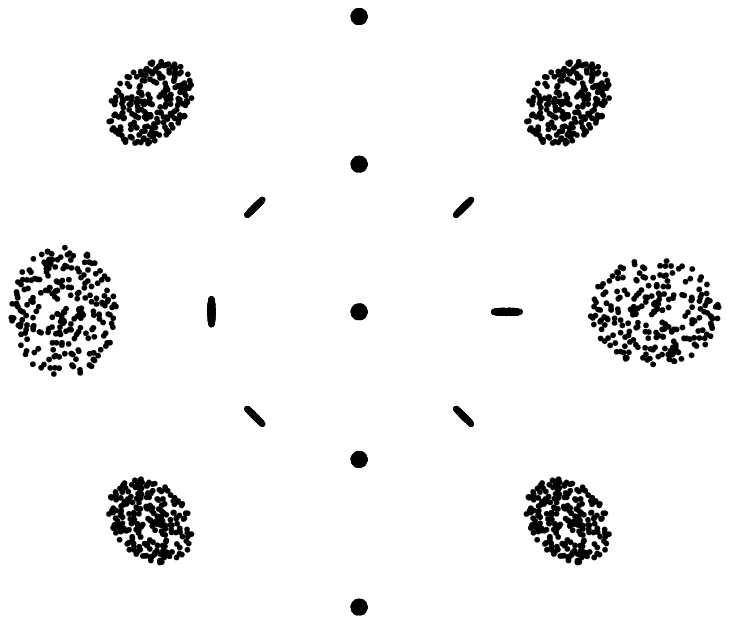}
\includegraphics{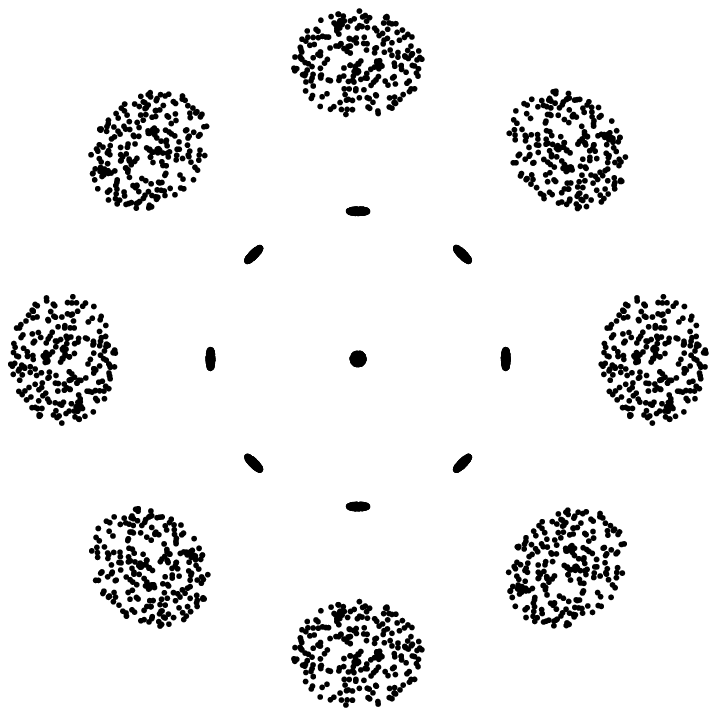} \includegraphics{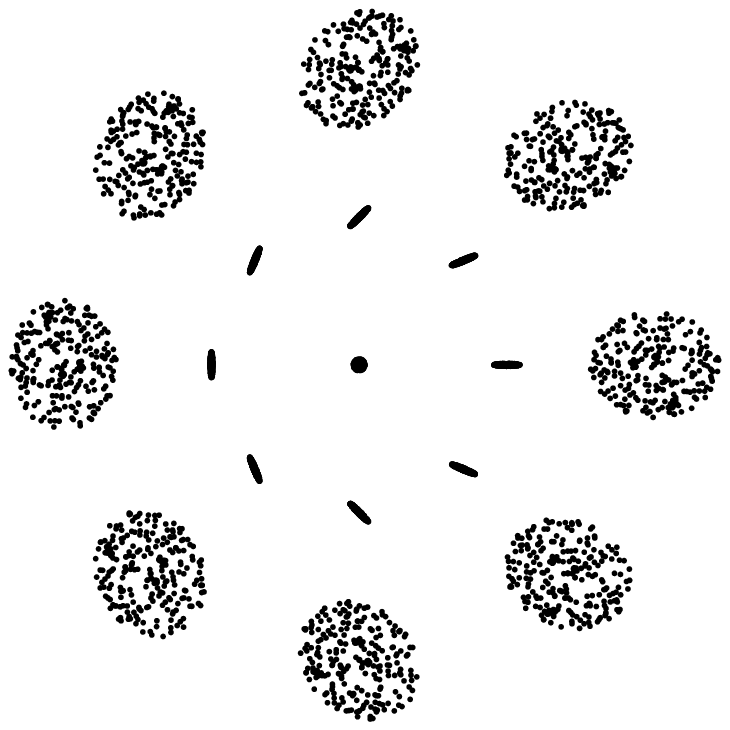}
\caption{Fifth-order astigmatism field patterns.  (a) 
  The ``$\sigma$'' astigmatism 
  pattern indicating a tilt and (or) decentering of a mirror
  along the x axis.  Note that the astigmatism is radially aligned.
  (b) Astigmatism pattern typical of an aligned telescope.  
 (c)   The ``$\rho$'' astigmatism 
  pattern indicating a tilt and (or) decentering of a mirror
  along the x axis.  Note that the azimuthal dependence on
  field angle is identical to that for third-order misalignment
  astigmatism.   A constant defocus has been added to all three
  panels to show the orientations
  of the aberrations.
  \label{fig:5th_astig_fp_s4}}
\end{figure}

The PSF for fifth-order astigmatism is identical to that of Seidel,
third-order astigmatism.  But the two misalignment aberration patterns
shown in Figure \ref{fig:5th_astig_fp_s4} are unlike third-order
misalignment astigmatism (see Figure \ref{fig:astig_s2}).  The $\rho$
misalignment pattern has the same azimuthal field dependence as for
third-order astigmatism, but has a different radial dependence.  The
$\sigma$ misalignment pattern has a different azimuthal field
dependence and a different radial dependence.  The ability to
distinguish between the two fifth-order astigmatism misalignment
patterns and the third-order astigmatism misalignment pattern
therefore depends critically upon the sampling of the field rather
than the wavefront.

\subsubsection{fifth-order coma: misalignment}

\begin{figure}[htbp]
\vspace{2.0 truein}
\includegraphics{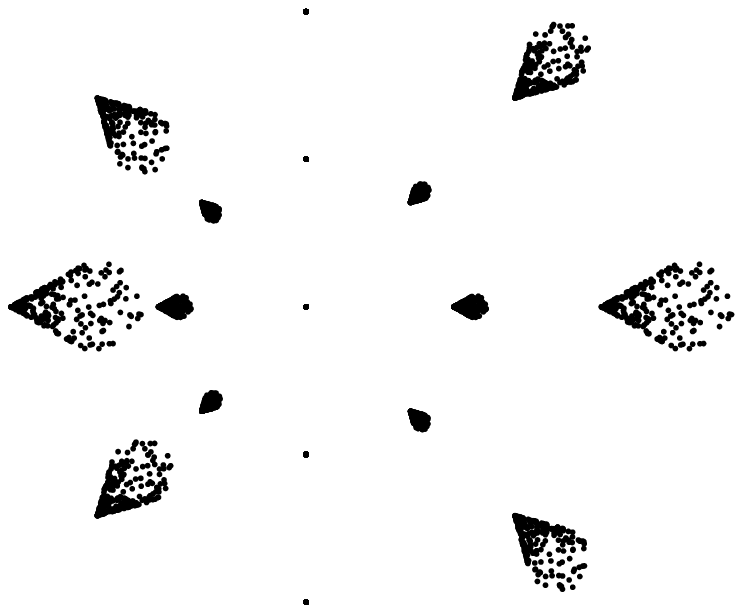}
\includegraphics{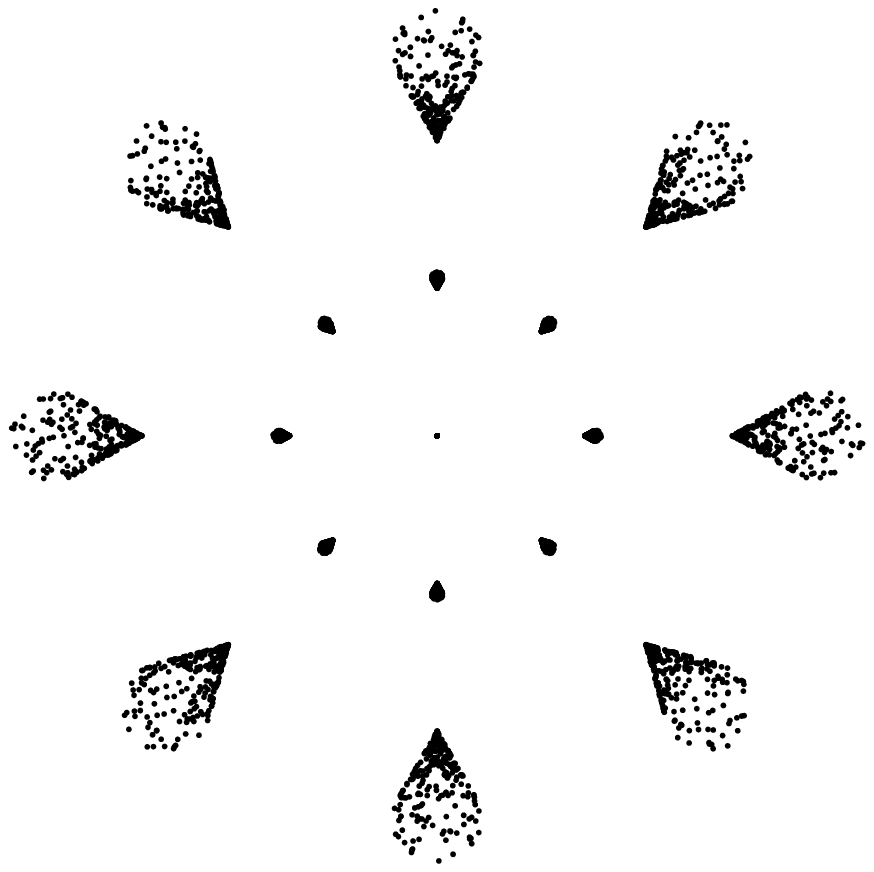}
\includegraphics{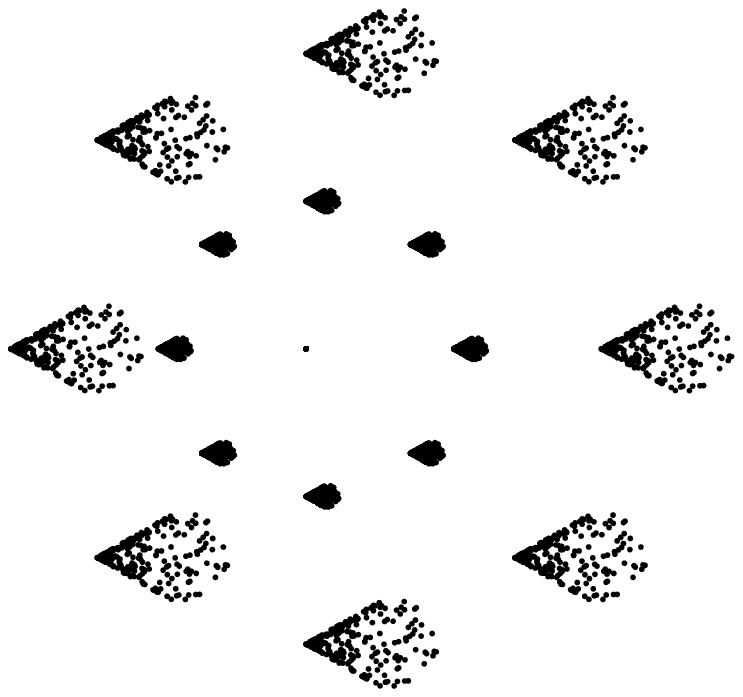}
\caption{Fifth-order coma field patterns.  (a) 
  The ``$\sigma$'' coma
  pattern indicating a tilt and (or) decentering of a mirror
  along the x axis.  Note that the coma is radially aligned.
  (b) Fifth-order coma
  characteristic of an aligned telescope  
  (c)   The ``$\rho$'' coma
  pattern indicating a tilt and (or) decentering of a mirror
  along the x axis.  
  Note that the azimuthal dependence on
  field angle is identical to that for third-order misalignment
  coma. \label{fig:5th_coma_fp_s4}}
\end{figure}

As with fifth-order astigmatism, the PSF for fifth-order coma is
identical to that of Seidel, third-order coma.  But the two
misalignment aberration patterns shown in Figure
\ref{fig:5th_coma_fp_s4} are unlike third-order misalignment coma (see
Figure \ref{fig:coma_s2}).  The $\rho$ misalignment pattern has the
same azimuthal field dependence as for third-order coma, but different
radial dependence.  The $\sigma$ misalignment pattern has a different
azimuthal field dependence and a different radial dependence.  As
with fifth-order astigmatism, the ability to distinguish between the
two fifth-order coma misalignment patterns and the third-order coma
misalignment pattern therefore depends critically upon the sampling of
the field rather than the wavefront.

\subsubsection{fifth-order defocus: misalignment}

As with fifth-order astigmatism and coma, the PSF for fifth-order
defocus is identical to that of curvature of field (which might
equally well be called third-order defocus).  But the misalignment
aberration patterns are not identical.  They have the same angular
dependence, but exhibit a different radial dependence.  Once again,
the ability to distinguish between the fifth-order defocus
misalignment pattern and the curvature of field misalignment pattern
depends critically upon the sampling of the field rather than the
wavefront.

\subsection{discussion of fifth-order misalignment aberration patterns}

The literature on fifth-order aberrations is limited for several
reasons.  First, by their very nature, they tend to be smaller than
the third-order (Seidel) aberrations.  Second, they are rather
cumbersome.  As a matter of course, ray tracing programs handle them
correctly, so as a matter of practice, they receive little attention.

But as we have seen in the previous section, fifth-order aberrations
may be needed to keep a three-mirror telescope aligned, and would
almost certainly be needed to keep a four-mirror telescope aligned.
Conversely, if one uses only the more commonly measured 
third-order aberrations, coma, astigmatism, and COF,  to keep a
telescope aligned, there are degenerate telescope configurations that
zero out these third-order misalignment aberrations and yet produce 
distortion and fifth-order
misalignment aberrations. It is therefore of some interest to
estimate which of these might be more or less substantial.  

An order of magnitude argument can be made by noting that the entries
in Table \ref{tab:5th_order} are homogeneous and of sixth order in the
sum of the exponents of the pupil radius and the field angle.  Both of
these are rendered dimensionless by the focal lengths.  We would argue
that it is the focal ratio of the fastest element that matters most in
such considerations.  Modern fast, wide field telescopes have primary
f-ratios approaching unity, while the field angles are rarely greater
than a tenth of a radian.  By this argument the terms at the top of
Table \ref{tab:5th_order} would be larger than those at the bottom. 
Third- and fifth-order aberrations for three specific optical systems: the 
LSST, HET corrector, and a Magellan-like telescope, are analyzed in the 
following subsections.

\subsection{Tessieres' models for the LSST}

The approach advocated here is quite similar to that adopted by Regis
Tessieres in an unpublished M.S. thesis \citeyearpar{Tessieres2003}
carried out at the University of Arizona.  \citeauthor{Tessieres2003}
analyzed the off-axis aberrations for two telescopes in terms of
patterns derived in Thompson's unpublished \citeyearpar{Thompson1980}
PhD thesis.  But instead of computing the amplitudes of the patterns
according to the principles set forth by \citeauthor{Thompson1980}, he
used ray-tracing software to produce wavefronts across the
field for various misalignments.  He then fit these wavefronts to the
expected patterns.

Of particular relevance for the present work, he analyzed an early
version of the Large Synoptic Survey Telescope (\citealt{Seppala2002};
henceforth LSST) in which the tertiary and primary were independent
(rather than fabricated from a single monolith, as with the ultimate
design).  He applied decenters and tilts to the secondary and
tertiary (and to the corrector assembly) and decomposed the computed
wavefronts into the third- and fifth-order aberration patterns.
Results of those calculations are given in Table \ref{tab:Tessieres},
which show the effects of tilts in the secondary and
tertiary.\footnote{\citeauthor{Tessieres2003}' nomenclature is similar
  to \citeauthor{Hopkins1950}', but each aberration is preceded by its
  field angle dependence.  Thus what we would called third-order
  misalignment astigmatism he calls field-linear astigmatism.}  Each
entry gives the amplitude of the aberration at the edge of the field
and at the edge of the pupil, in microns, for one degree of tilt.  
Up to factors of order unity the rms
spot size will be proportional to these.

The first impression one gets is that the third-order aberrations are
factors of 30-300 larger than the fifth-order aberrations.  A
consequence of this disparity is that, to the extent that the
fifth-order and third-order aberration patterns are correlated, a
small relative error in a measurement of a third-order aberration
pattern will produce a large relative error in the corresponding
fifth-order aberration pattern.  This bodes ill for using
fifth-order aberrations for telescope alignment.
\citeauthor{Tessieres2003}' calculations are themselves not entirely
immune from such errors, but \citeauthor{Tessieres2003} had the luxury
of measuring the wavefront with high precision at a large number of
points in the field.

There are several patterns for which \citeauthor{Tessieres2003} did
not report amplitudes -- fifth-order astigmatism-$\sigma$ and the
$\rho$ patterns for fifth-order coma and astigmatism for the secondary
mirror.  One suspects that the amplitudes for these were so small as
to be in the noise, but \citeauthor{Tessieres2003} is not explicit on
this point. 

 \citeauthor{Tessieres2003} {\it did} fit several patterns that vary
 quadratically with tilt and decenter.  We suspect that the
 coefficients for these would approach zero for successively smaller
 tilts and decenters.  Assuming iterative alignment correction these
 will play no role once the misalignments are small, and we have not
 included them in our discussion.

 \citeauthor{Tessieres2003} did not measure the misalignment patterns
 for either spherical aberration or curvature of field (defocus).
 These surely contributed to his figure of merit, and his alignment
 experiments might have converged more rapidly had he measured them.

Finally \citeauthor{Tessieres2003} did not measure distortion, which
might in principle be used to align a telescope.

\begin{deluxetable}{l l l l }
     \tablewidth{0pt}
     \tablecaption{Amplitude of misalignment patterns at edge of the field for
an early version of the LSST (in microns) for tilts of $1^{\circ}$.\label{tab:Tessieres}}
     \tablecolumns{4}
     \tablehead{
          \colhead{here} &
          \colhead{Tessieres} &
          \colhead{M2} &
          \colhead{M3} 
     }
     \startdata
third-order coma    &  constant coma 
          &  178
          &  186
          \\
third-order astigmatism   &  linear astigmatism 
          &  114
          &   36
         \\
          
coma-II   &  constant fifth-order coma
          &  2.4
          &  4.3
          \\
          
astigmatism-II & linear oblique spherical
          &  0.58
          &  0.17
         \\

trefoil   & quadratic elliptical coma
          & 0.41
          & 0.49
          \\
fifth-order coma-$\sigma$ &  quadratic coma\#2 
          & 0.90
          & 2.09
         \\

fifth-order coma-$\rho$  &  quadratic coma\#1 
          & $^a$    
          & 3.5
          \\
fifth-order astigmatism-$\rho$  &  cubic astigmatism\#1 
          & $^a$   
          & 1.72 
         \\ 
     \enddata
\tablenotetext{a}{\footnotesize no value given}
\end{deluxetable}

\subsection{Manuel's models for the HET corrector}

In another unpublished Ph.D. thesis, Anastacia Manuel
\citeyearpar{Manuel2009}, working at the University of Arizona,
carried out a ray-tracing misalignment analysis of a four-element
corrector for the Hobby-Ebberly telescope.

The emphasis was on identifying the combinations of motions of the
four elements that produced the largest aberrations using singular
value decomposition.  These modes sometimes involved more than one of
the patterns considered here.  Several of the larger modes were
associated with despacing and manifested themselves in symmetric
aberration patterns.

But of the misalignment patterns, the three largest were the
misalignments associated coma, curvature of field, and astigmatism,
all of which are third-order.  Next after that came linear
combinations of second coma and second astigmatism misalignment
patterns.  These modes produced a figure of merit (which we take to be
proportional to the wavefront error) a factor of $10^4$ smaller than
those for coma-I and a factor of 30 smaller than those for
astigmatism-I.  This would again suggest that measurements of
fifth-order aberrations may not contribute much to aligning the system
in question.

\subsection{modeled aberrations for a Magellan-like telescope}

In order to explore the feasibility of using distortion or the fifth-order 
aberrations for telescope alignment, we used Zemax\textsuperscript{\textregistered} to 
measure the third- and fifth-order aberrations of a Gregorian telescope 
adapted from the Magellan Baade and Clay telescopes.  The model telescope consists of three 
optical elements and a detector: a 6.4784m diameter aperture stop 
which sits 0.32512m in front of the primary mirror, a primary mirror with 
radius of curvature of -16.2553m and conic constant -1.00001, and a 
secondary mirror 9.72205m in front of the primary with radius of 
curvature 2.86282m and conic constant -0.63286.  The focal plane sits 
4.28026m behind the primary mirror.  We analyzed two telescope 
configurations: one with a 1mm decentered secondary mirror and one 
with a $1/4^\circ$ tilted secondary mirror.  The telescope specs appear in 
table \ref{tab:Magellanspecs}.

\begin{deluxetable}{l c c c c}
     \tablewidth{0pt}
     \tablecaption{Design specifications of the Magellan-like telescope.\label{tab:Magellanspecs}}
     \tablecolumns{5}
     \tablehead{
          \colhead{Surface} &
          \colhead{\begin{tabular}{c} Radius of \\  Curvature (m) \end{tabular} } &
          \colhead{Diameter\tablenotemark{a} (m)} &
          \colhead{Conic Constant} &
          \colhead{\begin{tabular}{c} Distance (m) to \\ Next Surface \end{tabular} }
     }
     \startdata
     object plane		& - 			& - 			& - 			& infinity \\
     aperture stop	& - 			& 6.4784		& - 			& 0.32512000 \\
     primary mirror	& -16.2553	& 6.4805  		& -1.00001	& -9.72205034 \\
     secondary mirror	& 2.86282		& 1.6134		& -0.63286	& 14.00230590 \\
     image plane		& - 			& - 			& - 			& -
     \enddata
\tablenotetext{a}{\footnotesize Diameter of the optic is larger than the diameter of the beam at the optic.}
\end{deluxetable}

\afterpage{\clearpage}

For each telescope configuration, the Zemax\textsuperscript{\textregistered} analysis tool
{\it Fringe Zernike} was used to compute the Zernike wavefront aberrations 
at the edge of the pupil for a total of 21 field points spanning the $x$ and $y$ 
axes of the field.  For the points along each telescope axis, the Zernikes 
corresponding to the pupil variations given in tables \ref{tab:3rd_order} and 
\ref{tab:5th_order} were then fit to the field patterns using 
a simple least squares algorithm.  
Measurement of the wavefront delay along two field axes was necessary 
to distinguish the effects of $\rho$ and $\sigma$ fifth-order astigmatism 
and also $\rho$ and $\sigma$ fifth-order coma.  For the other 
aberrations the second field axis provided a redundancy with which to 
verify the results.  All of those third-order patterns fitted redundantly are 
self-consistent to within $0.2\%$.  The fifth-order patterns are 
self-consistent to within $2\%$ with the exception of second coma for the 
tilted telescope and misalignment trefoil for the decentered telescope.  
The former is likely corrupted by higher order effects which will be 
discussed later, and the latter is consistent with zero.

We also used the Zemax\textsuperscript{\textregistered} analysis tool {\it Grid Distortion} to analyze 
third- and fifth-order distortion for both telescope configurations.  As with 
fifth-order coma and fifth-order astigmatism, we fit the distortion and 
fifth-order distortion patterns along two separate field axes in order to 
distinguish between the effects of the $\rho$ and $\sigma$ 
misalignment patterns.

The fitted aberrations for third- and fifth-order and the corresponding 
analytically computed aberrations for third-order appear in table \ref
{tab:Magellanish}.  The coefficients are given in microns of wavefront 
delay at the edge of the pupil for an image at the edge of a $1^\circ$ 
field, with the exception of distortion and fifth-order distortion which are 
given in arcseconds of image displacement at the edge of the field.

\begin{deluxetable}{l c c c }
     \tablewidth{0pt}
     \tabletypesize{\footnotesize} 
     \tablecaption{Amplitudes of aberration patterns for
a Magellan-like telescope.\label{tab:Magellanish}}
     \tablecolumns{4}
     \tablehead{
          \multirow{2}{*}{Aberration} &
          \multirow{2}{*}{Symmetric} &
          Misalignment ($\rho$,$\sigma$) &
          Misalignment ($\rho$,$\sigma$)  \\[-0.7ex]
           &
           &
          \scriptsize{Decentered 1mm} &
          \scriptsize{Tilted $1/4^\circ$} 
     }
     \startdata
\multicolumn{4}{c}{} \\[-4.5ex]
\multicolumn{4}{c}{Third-order} \\
\hline
\\ [-2.25ex]
spherical    &  -0.005	&	&
          \\
coma   &  -29.161	&-15.600	&108.527          
         \\          
astigmatism  &  129.592	&-0.479	&51.848        
         \\          
COF & -591.745	&7.631	&-72.386
         \\          
distortion\tablenotemark{a} & -7.908	&0.079	&-1.363
         \\          
          & & 0.098 	&-0.619\\
\multicolumn{4}{c}{Fifth-order} \\
\hline
	\\ [-2.25ex]
spherical-II & 0.000	&	&
         \\          
coma-II &  0.013	&0.587	&-4.087          
         \\          
astigmatism-II &  1.005	&-0.014	&0.031
         \\          
fifth-order spherical &  -0.592	&0.039	&-0.248         
         \\          
trefoil &  0.261	& -0.001	& 0.263
         \\          
fifth-order coma &  3.641	&-0.037	&-0.375
         \\          
         & & -0.055		&0.286
         \\          
fifth-order astigmatism &  -0.031	&0.001	&-0.143
         \\          
                  & & -0.003 	&0.280
         \\          
fifth-order defocus &  0.102	&0.002	&-0.168
         \\          
fifth-order distortion\tablenotemark{a} &  0.010	&0.000	&-0.003        
         \\          
	&	&	 0.024	&0.027\\
\hline
\multicolumn{4}{c}{Analytic third-order} \\
\hline
	\\ [-2.25ex]
spherical    &  -0.005	&	&
         \\          
coma   &  -29.180	&-15.625	&108.702          
         \\          
astigmatism  &  129.731	&-0.425	&51.485        
         \\          
COF & -591.864	&7.660	&-29.027
         \\          
distortion\tablenotemark{a} & -7.787	&0.078	&-1.369
         \\*         
          & 			& 0.112 	&-0.875
     \enddata
\tablenotetext{a}{\footnotesize Distortion and fifth-distortion are reported in arcseconds of displacement at the edge of a $1^\circ$ field.  All other values are reported in microns of wavefront delay at the edge of the pupil. }
\end{deluxetable}

\afterpage{\clearpage}

While the magnitude of misalignment aberrations are specific to 
telescope design, for the Magellan-type telescope modeled here, the 
fifth-order misalignment aberrations are never more than $4\%$ of the 
magnitude of the third-order misalignment coma aberration (and more 
often than not, less than $1\%$ of misalignment coma).  While the 
misalignment distortion terms are also small, $-1\farcs363$ and $-0\farcs619$ 
for $\rho$ and $\sigma$ distortion respectively in the tilted telescope, they 
are significant compared to the reported RMS error of the SDSS astrometric 
catalog, $0\farcs045 - 0\farcs075$ and $0\farcs100$ for stars with magnitudes up to 20 and 22 \citep{PierMunn2003}.  If the wavefront sensors on this 
hypothetical telescope sample 
the pupil well, second coma is the most easily measured fifth-order 
aberration with which to align the secondary.  If field sampling is preferable to 
pupil sampling and either a catalog exists for the field or one is willing to dither on the field \citep{Sudol2011}, distortion could be 
used for alignment.

In general, there is good agreement between the magnitudes of the third-order aberrations computed analytically using the formulae presented here and by Zemax\textsuperscript{\textregistered}.  By far the most significant discrepancy between the analytically computed aberrations and those measured by Zemax\textsuperscript{\textregistered} is for misalignment curvature of field for the tilted mirror.  This discrepancy could be due to an inadvertent focal plane tilt with the mirror tilt.  Zemax\textsuperscript{\textregistered}, however, reports that the focal plane is aligned with the primary mirror.

Interestingly, seventh-order effects become important for the measurement 
of the spherical and fifth spherical, second coma, and second 
spherical aberrations.  For each of these aberrations, 
the next order aberration intrinsic to the aligned telescope 
(seventh-order spherical, seventh-order second coma, and 
seventh-order second spherical) is of a 
similar magnitude or greater than the symmetric fifth-order aberration.  
Fortunately for telescope alignment, the corresponding seventh-order 
misalignment aberrations are smaller than the fifth-order misalignment 
aberrations.

\section{Wavefront Sensors: How Many, Where and to What Order?}
\label{sec:WFS}

\subsection{General considerations}

In \S \ref{sec:generic_patterns} and \S \ref{sec:generic_fifth-order} 
we showed that telescope misalignments produce a finite
and relatively small set of distinct aberration patterns, each of
which is characterized by a two-vector.  The degrees of freedom that
produce these patterns are likewise finite and small in number.

Suppose that we have an $N$-mirror telescope with $n$ wavefront
sensors distributed throughout the field each capable of measuring $m$
aberrations.  We take the position and pointing of the telescope to
be defined by one of the mirrors.  There remain $N-1$ tilts and $N-1$
decenters, each of which is described by a two-vector, that can produce 
our misalignment aberration patterns.  If the aberration patterns are
linearly independent, one would need to measure 
and correct $2(N-1)$ misalignment patterns to keep the
telescope aligned. 

\subsection{A naive approach to the 3-mirror telescope using trefoil}

For a 3-mirror telescope one would certainly measure the
coma, astigmatism, and curvature of field patterns, since these
produce the largest aberrations.  The choice of a fourth pattern is
much less obvious, but for the sake of the present exposition, we
choose trefoil (despite its relatively small amplitude compared to
several other fifth order aberrations).

Coma, astigmatism and trefoil each have two independent components on
the pupil: one that varies as $\cos \phi$, $\cos 2\phi$ and $\cos 3\phi$
respectively and another that varies as $\sin \phi$, $\sin 2\phi$ and
$\sin 3\phi$ respectively.  Thus, if the Seidel patterns and effects of despace 
errors and mirror 
deformations are well-constrained, one wavefront sensor alone will suffice
to measure the two-vectors for the misalignment aberration patterns 
associated with these.

By contrast, defocus is radially symmetric on the pupil, and each
wavefront sensor produces only one number.   Two wavefront sensors
suffice to measure the curvature of field pattern, provided that
they are not diametrically opposed.  

The situation is complicated by despace (piston) errors which produce
the symmetric aberration patterns characteristic of aligned but
incompletely corrected telescopes.  One would certainly want to
measure and correct the first-order defocus that this
produces, but this demands a third wavefront sensor to
distinguish between simple (uniform) defocus and curvature of field.

One possible arrangement of these three wavefront sensors would 
be to space the wavefront sensors
around the periphery of the field, where the effects of the
misalignments are largest.  An alternative would be to have one
wavefront sensor at the center of the field and two at the periphery,
but not co-linear.

As there are two degrees of freedom for despace errors, one must
measure a second field-symmetric pattern in addition to defocus.  The
three candidate patterns are coma, astigmatism and spherical
aberration.  Only two wavefront sensors are needed to
distinguish between the third-order misalignment coma and the symmetric
coma pattern characteristic of a telescope with despace errors.
The same holds true for astigmatism.  The three wavefront sensors
needed to distinguish uniform defocus and curvature of field therefore
provide information about both despace errors.  In addition, one
would obtain three measurements of spherical aberration, which at
third order is constant across the field.

Our naive scheme has one major strength and one major weakness.  The
strength is that the aberrations are low order on the pupil.
Wavefront sensors work by subdividing the pupil and making
measurements on those elements.  The more finely one subdivides the
pupil, the more light is needed.  Moreover, high-order aberrations are
likely to be small, so there is little danger of our low order
aberrations being corrupted by covariances with higher order
aberrations.  Finally, for the case of wavefront sensors based on
out-of-focus images, image overlap is less of a problem if one can
work with images that are more nearly in focus, which this method allows.

The substantial weakness in our scheme is the use of trefoil, which,
as we have noted, would appear to produce a substantially smaller
signal than some of the other fifth-order misalignment aberrations.

\subsection{alternatives to trefoil: second coma, astigmatism and 
spherical}

Tessieres' results, shown in Table 3, would suggest second coma as an
alternative to trefoil.  But as is evident if Figure 7, the point
spread functions for second coma and ordinary (third order) coma are
quite similar to each other.  And the misalignment patterns, shown in
Figures 2 and 8, are identical.  So one would need to sample the pupil
well in order to distinguish between the two.

Coma and second coma vary,
respectively, as $\rho^3 \cos\phi$ and $\rho^5 \cos\phi$.   We let 
\begin{equation}
C = \frac{
      \int^{r_o}_{r_i}(\rho^5) (\rho^3) \rho d \rho  
    }
   {   \left[\int^{r_o}_{r_i} (\rho^5) (\rho^5) \rho d \rho
      \int^{r_o}_{r_i} (\rho^3) (\rho^3) \rho d \rho           
      \right]^{\frac{1}{2}}                                                          
   }
\end{equation}
be a measure of the correlation of the two wavefronts, where we have
suppressed the azimuthal dependence and assumed that the
center of the pupil is obstructed out to some fraction $r_i$ of its total radius.
Taking $r_i = 0.5$ and $r_o =
1.0$ gives $C = 0.981$, indicating a very strong correlation.  By
contrast $C$ is identically zero for any two of defocus, coma,
astigmatism and trefoil.  

As a consequence of this strong correlation, one needs very accurate
wavefront sensing to keep the large third-order coma signal 
from corrupting the weaker second-coma signal.

The problem is exacerbated by the large central obscurations
associated with wide-field telescopes.  One must obtain at least
two radial samples in the range $0.5 < \rho < 1.0$.  For a Shack-Hartmann
sensor, this would mean at least 8 lenslets across the diameter of
the pupil.  For out-of-focus wavefront sensing this would mean defocussing
by at least 8 seeing disks.

The same argument holds for discriminating between ordinary (third
order) astigmatism and second astigmatism, but with a slightly smaller
correlation, $C = 0.978$.  And it holds again for discriminating
between curvature of field and second spherical aberration, again
with with $C = 0.978$.

\subsection{alternatives to trefoil: fifth-order coma and astigmatism}

The PSFs produced by the $\sigma$ and $\rho$ fifth-order misalignment
coma are indistinguishable from those produced by third-order
misalignment coma, but each of these produces substantially different
field patterns.  One can therefore hope to distinguish among them
using additional wavefront sensors.

Since each of these patterns is characterized by a two-vector, and since
each wavefront sensor gives only two numbers associated with the coma
at that point, a bare minimum of three wavefront sensors are needed
to distinguish among the three misalignment coma patterns.

Of the two alternative arrangements described in our naive approach,
the one with all three wavefront sensors on the periphery fails to
discriminate between the fifth-order $\rho$-coma pattern and the third-order 
misalignment coma pattern.  By contrast the arrangement with one
at the center and two non-colinear on the periphery {\it does}
successfully discriminate among all three patterns.

One can still use only peripheral wavefront sensors as long as one was
careful to include the contributions of misalignments to both the
third-order coma pattern and fifth-order $\rho$-pattern.  In solving
for the misalignments, one might iterate, ignoring $\rho$-coma on the
first iteration and then subtracting it off on subsequent iterations.
But this would preclude counting $\rho$-coma as one of the $2(N-1)$
needed patterns.

The same arguments hold for the three misalignment astigmatism
patterns as well.  A strictly peripheral arrangement of wavefront
cannot distinguish between the third-order misalignment astigmatism
pattern and the fifth-order $\rho$ pattern.

\subsection{alternatives to fifth-order: pointing and distortion}
One could alternately choose to neglect fifth-order aberrations and instead
use pointing or distortion to maintain alignment in a three mirror telescope.  

Tilt (or pointing) is one of only two first-order aberrations.  As its name implies, 
pointing uniformly shifts the location of every object in the field. It does not alter 
an image's point spread function and so it cannot be measured with a wavefront 
sensor unless the position of that wavefront sensor in the image plane is 
accurately known. In a paper on the Advanced Solar Telescope, 
\citet{ManuelBurge2009} suggest that one might use pointing to constrain telescope alignment.  This method would require accurate measurement of the position of the detector with respect to one of the mirrors.

In the same way that spherical aberration might equally well have been
called ``second defocus'' -- the two have same dependence upon pupil
azimuth -- distortion might equally well have been called
``second tilt.''   For a field with a pre-existing astrometric catalog, misalignment distortion can be measured directly from the science 
data or using two wavefront sensors whose positions are accurately determined.  In the absence of a pre-existing catalog, one might also solve for the misalignment distortion patterns by comparing overlapping fields, as suggested by \citet{Sudol2011}.  The distortion terms would then constrain four degrees of freedom of telescope misalignments.

\subsection{An improved approach to the 3-mirror telescope}

Both wavefront sensor arrangements adopted for our naive 
approach to the 3-mirror telescope suffice to keep it aligned,
but not for the same reasons.  For the arrangement with three peripheral
wavefront sensors, either the $\sigma$-coma and $\sigma$-astigmatism
is likely to give a stronger alignment signal than the trefoil pattern.

Alternatively, for the arrangement with one wavefront sensor at the
center and two non-colinear sensors on the periphery, the $\rho$-coma
and $\rho$-astigmatism can also be used.  This would seem to give
sufficient information to keep a 4-mirror telescope aligned, or
a 3-mirror telescope with a corrector.  

It should be noted, however, that if one measures all three
misalignment coma patterns, one must add a fourth wavefront sensor if
one also wants to measure the symmetric coma pattern due to despace
errors.  The same holds true for astigmatism.

Table \ref{tab:WFS} is a summary of misalignment and piston errors, and the 
number of wavefront sensors required to constrain them.  The aberrations are
divided into those measurable with very low order wavefront sensing, and 
those which require greater sampling of the pupil. Patterns that
are degenerate in a single wavefront sensor and therefore require separate 
wavefront sensors to distinguish among them are grouped together.  
The P's and M's show the patterns one would measure with a minimal 
system of three wavefront sensors all at the same radius on the periphery:
P denotes those patterns which can be used to constrain piston; M 
misalignments.  Six independent misalignment patterns are 
measured, more than sufficient for a three mirror telescope and barely 
sufficient for a four mirror telescope.

\begin{deluxetable}{@{\extracolsep{\fill}} l r}
     \tablewidth{0.75\textwidth}
     \tablecaption{Wavefront sensing summary.\label{tab:WFS}}
     \tablecolumns{2} 
     \tablehead{
              \multicolumn{2}{l}{VERY LOW ORDER WAVEFRONT SENSING}
		}
     \startdata
	first-order symmetric defocus 			& 	P	\\
	third-order symmetric defocus (COF) 	& 	   	\\
	third-order misalignment defocus (COF)	&	M\tablenotemark{a}	\\[1.5ex]
	third-order symmetric astigmatism		&	P	\\
	third order misalignment astigmatism	&	M	\\
	fifth-order misalignment astigmatism-$\sigma$&	M	\\
	fifth-order misalignment astigmatism-$\rho$	& 	\tablenotemark{b}	\\[1.5ex]
	third-order symmetric coma			&	P	\\
	third order misalignment coma			&	M	\\
	fifth-order misalignment coma-$\sigma$	&	M	\\
	fifth-order misalignment coma-$\rho$		&	\tablenotemark{b}	\\[1.5ex]
	fifth-order misalignment trefoil			&	M	\\[1.25ex]
	\hline
	\hline
	\\[-2.0ex]
	\multicolumn{2}{l}{HIGHER ORDER WAVEFRONT SENSING}\\[0.5ex]
	\hline
	\\[-2.25ex]
	third-order symmetric spherical			& 		\\[0.75ex]
	fifth-order misalignment second-astigmatism	& 	\\[0.75ex]
	fifth-order misalignment second-coma		& 	\\[0.75ex]	
	fifth-order misalignment spherical			& 	
     \enddata
     \tablenotetext{a}{\footnotesize Third-order misalignment defocus alone requires 
         two wavefront sensors to fully constrain, even in the absence of 
         first- or third-order symmetric defocus.} 
     \tablenotetext{b}{\footnotesize If only three wavefront sensors are used, the four 
     	astigmatism patterns and the four coma patterns are degenerate.  But 
	if the misalignment can be determined from the remaining patterns, 
	one of these can be computed from the alignment solution.    
	For rho-coma this degeneracy is broken by placing a
	wavefront sensor at the center of the field.  However, the
	astigmatism degeneracy cannot be broken this way.}  
\end{deluxetable}

\section{Complications}
\label{sec:complications}

\subsection{mirror deformation}

We have until now treated the mirrors of a telescope as rigid bodies.
But under the influence of changing gravitational and thermal
stresses, the mirror surfaces deform and influence the wavefront.
Deformations of the mirrors can be expanded in terms of Zernike
polynomials, but it is more efficient to expand them in terms
of their elastic bending modes \citep{Noethe1991, MartinCallahan1998, 
SchechterBurley2003}.   

If the stop is coincident with one of the mirrors, then the
deformations of that mirror have the same effect on the wavefront at
every point in the field.  But if not, deformations of a mirror will
project onto different parts of the pupil at different points in the
field.  Applying the same mirror deformation pattern to different
mirrors will produce aberrated wavefronts that are offset
with respect to each other. One must therefore sample the wavefront
at relatively high density to ascertain which element is deformed.

Among the fifth-order aberrations, second-coma, second-astigmatism
and fifth-order spherical likewise require good sampling of
the pupil to distinguish their misalignment patterns from those
of third order aberrations.   The accuracy with which one might
determine these fifth-order aberrations is diminished by the need to use the 
same information to determine the mirror deformations.  

This would argue for using two kinds of wavefront sensors: low-order
sensors, using only coma, astigmatism and defocus to correct the
rigid body motions, and high-order sensors to correct for mirror
deformations.  These latter sensors might be run at lower cadence than the 
former, except perhaps immediately following a large change in telescope 
pointing.

In such a scheme one might still measure fifth-order misalignment coma and fifth-order misalignment astigmatism with the low-order wavefront sensors, provided 
that there are a sufficient number of such sensors.  If the 
telescope is then properly aligned the second-coma and
second-astigmatism misalignment patterns will be zeroed out on the
high-order wavefront sensors, giving a cleaner measurement of the
mirror deformations in the high-order sensors.  

\subsection{central obscuration}

Many wide field telescope designs have a large central obscuration.
As mentioned in \S \ref{sec:WFS}, this would make it more difficult to 
distinguish between second coma
and second astigmatism on the one hand and ordinary coma and
astigmatism on the other, which in third-order have the same field
pattern.  The smaller range of pupil radii would increase correlated
errors.

\subsection{focal plane misalignments}

We have until this point avoided the question of how one knows the
position of one's wavefront sensors or one's field with respect to one or 
another of a telescope's mirrors.  As noted in the discussion of pointing
in the previous subsection, the position and tilt of the focal plane or wavefront 
sensors might be determined mechanically or interferometrically.  But if 
not, and if one wishes to correct for
focal plane misalignments, one must measure two additional aberration 
patterns, one for the tilt of the focal plane, and one for the decenter.

A tilted detector produces a field pattern identical to misalignment curvature 
of field.  However, if that pattern is used to correct
the detector tilt, it cannot also be used to keep the mirrors
aligned.

A decentered focal plane creates no aberrations of its own, but as is 
discussed in \S\S \ref{sec:aligning_2.5-mirror}, it can cause Seidel or 
fifth-order aligned aberrations to masquerade as misalignment aberrations.  
Measuring telescope aberrations relative to an incorrect field center will 
produce spurious misalignment aberrations even if the telescope is 
otherwise aligned. 

One can choose to fit for this focal plane decenter by measuring an 
additional field pattern.  If it is not feasible to measure an additional field 
pattern and one assumes that the center of the detector is the center of 
pointing, he can still completely null the 
measured misalignment coma, astigmatism and curvature of field by tilting 
and decentering one of the mirrors and tilting the focal plane.  Compensating 
for the detector decenter with mirror misalignments can put the telescope into 
a state of `benign misalignment', where only the smaller aberrations patterns: 
misalignment distortion, the fifth-order misalignment aberrations, and the
aberrations which vary as the square of misalignments are present.

\subsection{transmitting correctors}

In his treatment of the LSST, \citet{Tessieres2003} studies the effects of
tilts and decenters on a focal plane assembly consisting of
a multi-element corrector and the focal plane array.  Two additional
patterns are needed to keep this assembly aligned.  Interestingly,
the fifth-order aberration patterns produced by these tilts and
decenters are larger, compared to the third-order aberration
patterns, than for the secondary and tertiary mirrors.

\section{Using and not using misalignment aberration patterns}
\label{sec:to_use_or_not_to_use}

At least three of the generic misalignment aberration patterns
described in the previous sections are currently used to align
wide-field telescopes, and there may soon be reason to use more of
them.  Ignoring for the present, the question of whether or not to use
distortion\footnote{We take the view what can be measured can hurt
  you.}  we imagine here an $N$-mirror telescope with $n$ wavefront
sensors distributed throughout the field each capable of measuring $m$
aberrations.  One must determine $N-1$ tilts and $N-1$ decenters, each
of which is described by a two-vector.

\subsection{independent analysis of the wavefront sensors}

The most straightforward and transparent approach is to analyze each
wavefront sensor separately, determining the coefficients of the
aberrations at each of $m$ points in the field.  One then fits the
these coefficients to a linear combinations of the misalignment
aberration patterns described in the previous section and finally fits
the amplitudes of the misalignment patterns (assuming one has measured
more patterns than one actually needs).\footnote{It would not be
  surprising if two or more patterns were produced by the same, or
  nearly the same combination of tilts and decenters.}

A complication of this approach is that at each step the
quantities for which one is fitting may be correlated with each other.
Aberration coefficients will be correlated, pattern amplitudes will be
correlated and tilts and decenters will be correlated.  Under such
circumstances one must be careful to fit for all correlated
quantities; otherwise one runs the risk of introducing systematic
errors in the quantities for which one does fit.  

\subsection{simultaneous fit of all wavefront sensors for pattern amplitudes}

Instead of measuring the aberrations at each point in the field, one
might fit the data for all $n$ wavefront sensors simultaneously to
determine the amplitudes of the misalignment aberration patterns.
This has the advantage of eliminating the problem of correlated
aberrations at each point, but unmodelled aberrations may then lead to
systematic errors.  

\subsection{simultaneous fit of all wavefront sensors for tilts and decenters}

One might also fit directly for the tilts and decenters,
short-circuiting the misalignment aberration patterns except for using
them to turn tilts and decenters into predicted wavefronts that are
then compared with observed wavefronts.  This reduces the number of
parameters for which one fits and implicitly accounts for the
correlations of aberration pattern amplitudes.  Rather than
fitting for poorly determined amplitudes  of fifth-order aberration
patterns, all of the fifth-order dependence is attributed
to a smaller number of tilts and decenters, which are strongly
constrained by the third-order aberration patterns.

\subsection{forget about misalignment patterns}

Finally one might dispense entirely with the decomposition of the
wavefront into specific aberrations and instead use ray tracing to
determine how it varies at each measured point.  While this obviates
the need for misalignment patterns, it sacrifices all understanding of
why one might need $n$ wavefront sensors with enough resolution to
measure $m$ aberrations.

\section{Summary}

We have derived and illustrated the generic third-order aberration
patterns that arise when the axial symmetry of a telescope is broken
by small misalignments of optical elements.  There are five patterns:
one each for coma, astigmatism and curvature of field and two for
distortion.  Each of these misalignment patterns is characterized by
an associated two-dimensional vector.  These two-vectors are in turn
linear combinations of the tilt and decenter vectors of the individual
optical elements.  

For an $N$-mirror telescope, $2(N-1)$ patterns must be measured to
keep the telescope aligned.  For $N=3$, as in a three mirror
anastigmat, there is a two-dimensional ``subspace of benign
misalignment'' over which the misalignment patterns for third-order
coma, astigmatism and curvature of field are identically zero.  If
pre-existing astrometry is available, one or both of the distortion
patterns may be used to keep the telescope aligned.  Alternatively,
one might measure one of the fifth-order misalignment patterns.

We have illustrated the generic fifth-order misalignment patterns that
arise from small misalignments.  These are relatively insensitive to
misalignments and may be of little use in telescope alignment.  One would
appear to be driven back to using distortion, or alternatively,
pointing.

\medskip
\noindent
\acknowledgments
{\it Acknowledgements:} We gratefully acknowledge helpful and thought
provoking conversations and communications with Michael Jarvis, Don
Phillion, Lothar Noethe, Stephen Shectman and Tony Tyson.  We also thank
the referee, Brian McLeod, for his extensive comments.  This work was 
supported by the National Science Foundation through a Graduate 
Research Fellowship to Rebecca Sobel Levinson

\appendix
\section{Coma}

\citet{Schroeder1987}, calculates the the coma pattern cause by tilting and
decentering the secondary of a two mirror telescope by amounts $\alpha$ and $\ell$.  

\begin{align}
\label{eq:Schroeder_Coma}
     G & = B_2 \rho^3 \sin \phi \\ 
\nonumber     B_2 & = B_2(cen) + \frac{1}{R_2^2} \left[ \frac{l}{R_2} \left[K_2 - \left(\frac{m_2+1}{m_2-1} \right) \right] - \alpha\left(\frac{m_2+1}{m_2-1} \right) \right] \\ 
\nonumber     B_2(cen) &= \frac{\theta}{R_1^2} - \frac{W \theta}{R_2^2} \left[ \left(\frac{m_2+1}{m_2-1} \right) \left( \frac{1}{W} - \frac{1}{R_2} \right) + \frac{K_2}{R_2} \right] 
\end{align}

\noindent where $\rho$ and $\phi$ are the radial and angular coordinates on the pupil and $\theta$ is the radial coordinate of the image in the field.  The quantities  $R_i$, $K_i$ and $m_i$ are all mirror properties: the radius of curvature, conic constant, and magnification of a mirror, where the subscript denotes which mirror is being addressed. The index of refraction has been set equal to 1 for the primary mirror and -1 for the secondary mirror.  Finally, $W$ is the distance from the primary mirror to the secondary.    

$B_2(cen)$ is the coma of an aligned two mirror telescope. For the sake of simplicity the tilt and decenter of the secondary from the primary, and the object displacement from the optical axis have been taken lie along the $y$ axis in Schroeder's analysis.  

\citeauthor{Schroeder1987}'s equations can be generalized for an object displaced from the optical axis in an arbitrary direction.  The chief ray for the object is given by $\vec \sigma$ with radial and angular components $\sigma$ and $\theta$.  The equations can be further generalized to allow decentering and tilting of the secondary mirror in arbitrary directions $\vec \ell$ and $\vec \alpha$ with radial components $\ell$ and $\alpha$ and angular components $\phi_{\ell}$ and $\phi_{\alpha}$.  The equations become:

\begin{equation}
\label{eq:transition1_coma}
     G = B_{2x} \rho^3 \cos \phi + B_{2y} \rho^3 \sin \phi \\  
\end{equation}

\noindent where the field dependences are given by
\begin{align}
\label{eq:transition2_coma}
     B_{2x} & = G^{coma}_{Seidel}\sigma\cos\theta + G^{coma}_{decenter}\ell\cos\phi_{\ell} + G^{coma}_{tilt} \alpha\cos\phi_{\alpha}\\
\nonumber     B_{2y} & = G^{coma}_{Seidel}\sigma\sin\theta  + G^{coma}_{decenter}\ell\sin \phi_{\ell} + G^{coma}_{tilt}\alpha\sin \phi_{\alpha}\\
\nonumber     G^{coma}_{Seidel} &= \frac{1}{R_1^2} - \frac{W}{R_2^2} \left[ \left(\frac{m_2+1}{m_2-1} \right) \left( \frac{1}{W} - \frac{1}{R_2} \right) + \frac{K_2}{R_2} \right] \\
\nonumber     G^{coma}_{decenter}  & = \frac{1}{R_2^3} \left[K_2 - \left( \frac{m+1}{m-1} \right) \right] \\
\nonumber     G^{coma}_{tilt}   & = - \frac{1}{R_2^3} \left(\frac{m+1}{m-1} \right)
\end{align}

Consolidating the above equations yields

\begin{equation}
\label{eq:section2_coma}
G^{coma} = G^{coma}_{decenter}     \sigma  \rho^3\cos(\phi - \theta)
        +  G^{coma}_{decenter}   \ell   \rho^3\cos(\phi - \phi_\ell)
        +  G^{coma}_{tilt}      \alpha  \rho^3\cos(\phi - \phi_\alpha)
\end{equation}

\noindent which is the form for the coma aberration given in \S \ref{sec:generic_patterns}

\section{Astigmatism}

\citet{McLeod1996} follows the notation of \citeauthor{Schroeder1987} and calculates the astigmatism pattern for a two mirror telescope for which the secondary mirror has been aligned (decentered) to null the field constant coma pattern.
McLeod gives the form of the remaining astigmatism patterns as

\begin{align}
\label{eq:McLeod_astigmatism}
     & W = Z_4 \rho^2 \cos 2\phi +   Z_5 \rho^2 \sin 2\phi \\  
\nonumber     Z_4 &= B_0 (\theta_x^2 - \theta_y^2) + B_1(\theta_x \alpha_x - \theta_y \alpha_y) + B_2(\alpha_x^2 -\alpha_y^2)\\
\nonumber      Z_5 &= 2 B_0(\theta_x \theta_y) + B_1(\theta_x \alpha_y + \theta_y \alpha_x) + 2 B_2(\alpha_x \alpha_y)\\ 
\nonumber      & B_0 = A_0^p r_p^2 - A_0^s r_s^2\\
\nonumber      & B_1 = -r_s^2 \left[ 2A_0^s + (W+d_n) A_1^s \right] \\
\nonumber      & B_2 = -r_s^2 \left[ A_0^s + (W+d_n) A_1^s + (W+d_n)^2 A_2^s \right] \\
\nonumber      A_0 &= \frac{W^2}{2 R} \left[ \frac{K}{R^2} + \left( \frac{1}{W} - \frac{1}{R} \right)^2 \right] \\
\nonumber      A_1 &= \frac{W}{R^2} \left[ \frac{1}{W} - \frac{K + 1}{R} \right] \\
\nonumber      A_2 &= \frac{K+1}{2R^3}
\end{align}

\noindent where $\rho$ and $\phi$ are the (normalized) radial and angular coordinates on the pupil and $\theta_x$ and $\theta_y$ are the cartesian coordinates of the image in the field.  The quantities  $R_i$, $K_i$ and $m_i$ are again all mirror properties: the radius of curvature, conic constant, and magnification of a mirror, where the subscript denotes which mirror is being addressed, the primary $p$, or the secondary $s$.  $W$ is the distance from the primary mirror to the secondary, and $\alpha_x$ and $\alpha_y$ are the tilts of the secondary mirror with respect to the primary mirror.  The indices of refraction preceding the primary and secondary mirrors have been set to 1 and -1 respectively. 

\citeauthor{McLeod1996} additionally uses two constants that are not present in \citeauthor{Schroeder1987}'s analysis of the coma pattern: $r_i$, which is the marginal ray height at the optic, and $d_n$, which denotes the position of the coma free point.  
As \citeauthor{McLeod1996} nulled the misalignment coma prior to analyzing the astigmatism, the decenter of the system $\vec \ell$ is given by $\vec \ell = d_n \vec \alpha$.  \citeauthor{McLeod1996} chose to express the astigmatism only in terms of the tilt of the secondary mirror, though he could have equivalently expressed the astigmatism only in terms of the decenter or as a combination of the two terms.  For greater transparency of the field patterns caused by both decenters and tilts, we here decouple the decenter and tilt terms in \citeauthor{McLeod1996}'s equations, removing the variable $d_n$.  We also de-normalize the pupil coordinates and remove the terms which vary as the square of the misalignment.

\begin{align}
\label{eq:transition1_astig}
     & W = Z_4 \rho^2 \cos 2\phi +   Z_5 \rho^2 \sin 2\phi \\  
\nonumber     Z_4 &= B_0 (\theta_x^2 - \theta_y^2) + B_{1 decenter}(\theta_x \ell_x - \theta_y \ell_y) + B_{1 tilt}(\theta_x \alpha_x - \theta_y \alpha_y)\\     
\nonumber     Z_5 &= 2 B_0(\theta_x \theta_y) + B_{1 decenter}(\theta_x \ell_y + \theta_y \ell_x) + B_{1 tilt}(\theta_x \alpha_y + \theta_y \alpha_x) \\ 
\nonumber     & B_0 = A_0^p r_p^2 - A_0^s r_s^2\\
\nonumber     & B_{1 decenter} =  2A_0^s + W A_1^s\\
\nonumber     & B_{1 tilt} = A_1^s \\
\nonumber     A_0 &= \frac{W^2}{2 R} \left[ \frac{K}{R^2} + \left( \frac{1}{W} - \frac{1}{R} \right)^2 \right] \\
\nonumber     A_1 &= \frac{W}{R^2} \left[ \frac{1}{W} - \frac{K + 1}{R} \right] 
\end{align}

By converting the field variables to polar coordinates $\sigma$ and $\phi_\sigma$ and similarly converting the misalignment variables to polar form $\ell$, $\phi_\ell$, $\alpha$, $\phi_\alpha$, the expression for the wavefront delay yields the form given in \S \ref{sec:generic_patterns},
 
\begin{equation}
\label{eq:section2_astig}
G^{astig} = G^{astig}_{Seidel}     \sigma^2 \rho^2\cos2(\phi - \theta)
        +  G^{astig}_{decenter}   \sigma\ell   \rho^2\cos(2\phi - \theta - \phi_\ell)
        +  G^{astig}_{tilt}       \sigma\alpha  \rho^2\cos(2\phi - \theta - \phi_\alpha)
\end {equation}

\noindent where $B_0$, $B_{1 decenter}$ and $B_{1 tilt}$ are equal to $G^{astig}_{Seidel}$, $G^{astig}_{decenter}$ and $G^{astig}_{tilt}$.  \footnote{The form of the astigmatism field pattern holds even for telescopes which have not been aligned to null the misalignment coma pattern.  The expanded coefficients for the wavefront delay of a randomly misaligned telescope can be found in Table \ref{tab:3rd_order}.}

\section{Curvature of Field}

The misalignment patterns for curvature of field are less well-explored in the literature.  \citet{Thompson2005} presents forms for the patterns, and we re-derive them here from the despaces and misalignments of a single mirror from the pupil.  

As stated in \S \ref{subsec:generalization}, the contribution of a single mirror to the wavefront delay of the ray which strikes the mirror at position $\vec\varpi$ and makes an angle $\vec\psi$ with the axis of the mirror is given by:

\begin{align}
\label{eq:section2_3rd_order_surface}
G^{3rd} &= W_{040}(\frac{\vec{\varpi}}{R} \cdot
\frac{\vec{\varpi}}{R}) (\frac{\vec{\varpi}}{R} \cdot
\frac{\vec{\varpi}}{R}) + W_{131}(\vec{\psi} \cdot
\frac{\vec{\varpi}}{R}) (\frac{\vec{\varpi}}{R} \cdot
\frac{\vec{\varpi}}{R}) \\ \nonumber & + W_{222}(\vec{\psi} \cdot
\frac{\vec{\varpi}}{R})(\vec{\psi} \cdot \frac{\vec{\varpi}}{R}) +
W_{220}(\vec{\psi} \cdot \vec{\psi}) (\frac{\vec{\varpi}}{R} \cdot
\frac{\vec{\varpi}}{R}) \\ \nonumber & + W_{311}(\vec{\psi} \cdot
\vec{\psi})(\vec{\psi} \cdot \frac{\vec{\varpi}}{R})
\end{align}

\noindent 
where $W_{040}$ is the spherical aberration coefficient, $W_{131}$ is
the coma coefficient, $W_{222}$ is the astigmatism coefficient,
$W_{220}$ is the curvature of field coefficient, and $W_{311}$ is the 
distortion coefficient.   These aberration
coefficients depend only on the curvature of the mirror, $R$, the conic
constant of the mirror, $K$, the magnification of the mirror, $m$, the 
index of refraction of the air preceding the mirror, $n$ and
the position of the object for the mirror, $s$. 

The transformations from pupil
coordinates $\vec\rho$ and $\vec \sigma$ to mirror coordinates for a
mirror despaced by an amount $W$ and decentered and tilted by
$\vec{l}$ and $\vec{\alpha}$ are

\begin{align}
\label{eq:section2_surface2pupila}
\vec{\psi} &=(1-\frac{W}{s})\vec{\sigma} - (\vec{\alpha} + \frac{\vec{l}}{s}) \\
\label{eq:section2_surface2pupilb}
\vec{\varpi} &= (\vec{\rho} - W \vec{\psi}) - \vec{l}.
\end{align}

Expanding the wavefront delay caused by a single mirror in terms of the pupil coordinates and keeping only those terms which vary as $\rho^2$ on the pupil\footnote{The astigmatic component of the wavefront additionally has terms which vary as $\rho^2$ on the pupil.  We exclude these astigmatic terms in the analysis of COF} and vary linearly with the misalignments or less, we find an expression for the curvature of field wavefront delay added by a single offset mirror.

\begin{align}
\label{eq:Me_cof}
     G_i^{COF} &= \left[\frac{1}{2}\left(\frac{W}{s} -1 \right)^2W_{220} +  \frac{W}{R}\left(\frac{W}{s} -1 \right)W_{131} + 2\frac{W^2}{R^2} W_{040} \right] (\frac{\vec{\rho}}{R}\cdot\frac{\vec{\rho}}{R})(\vec{\sigma}\cdot\vec{\sigma}) \\
\nonumber     & \qquad + \left[ 2\left(\frac{R}{s}\right)\left(\frac{W}{s} + 1 \right)W_{220} + \left(\frac{2W}{s} + 1 \right) W_{131} + 4\frac{W}{R}W_{040}\right](\frac{\vec{\rho}}{R}\cdot\frac{\vec{\rho}}{R})(\vec{\sigma} \cdot \frac{\vec{l}}{R}) \\
\nonumber     & \qquad + \left[2\left(\frac{W}{s} +1 \right)W_{220} + \frac{W}{R} W_{131} \right] (\frac{\vec{\rho}}{R}\cdot\frac{\vec{\rho}}{R})(\vec{\sigma} \cdot \vec{\alpha})
\end{align}

For a two mirror telescope, the primary and the secondary both contribute to the wavefront delay.  As the primary is neither despaced nor misaligned from the pupil, the form of its contribution to the wavefront delay is simplified; notably, the primary mirror only contributes to the Seidel aberration.  The secondary mirror is despaced and possibly misaligned from its pupil and therefore contributes to the decenter and tilt terms as well as the Seidel pattern.  Combining the effects of the primary and secondary mirror yields:

\begin{align}
\label{eq:transition1_cof}
G^{COF} &= G^{COF}_{Seidel}  (\vec \sigma \cdot \vec \sigma)(\vec \rho \cdot \vec \rho)
       + G^{COF}_{decenter} (\vec \sigma \cdot \vec \ell)(\vec \rho \cdot \vec \rho)
       + G^{COF}_{tilt}  (\vec \sigma \cdot \vec \alpha)(\vec \rho \cdot \vec \rho )\\
\nonumber       & G^{COF}_{Seidel} = R_1^2 W_{220_1} + R_2^2\left[\frac{1}{2}\left(\frac{W}{s_2} -1 \right)^2W_{220_2} +  \frac{W}{R_2}\left(\frac{W}{s_2} -1 \right)W_{131_2} + 2\frac{W^2}{R_2^2} W_{040_2} \right] \\
\nonumber       & G^{COF}_{decenter} = -R_2^3  \left[ 2\left(\frac{R_2}{s_2}\right)\left(\frac{W}{s_2} + 1 \right)W_{220_2} + \left(\frac{2W}{s_2} + 1 \right) W_{131_2} + 4\frac{W}{R_2}W_{040_2}\right]\\
\nonumber       & G^{COF}_{tilt} =  - R_2^2 \left[2\left(\frac{W}{s_2} +1 \right)W_{220_2} + \frac{W}{R_2} W_{131_2} \right]
\end{align}

\noindent which is the form for the COF wavefront delay that appears in \S \ref{sec:generic_patterns}.

\section{Distortion}

The distortion field patterns can additionally be derived from the wavefront delay caused by a single optic.  Again expanding equation \eqref{eq:section2_3rd_order_surface} in terms of pupil coordinates, but now retaining only those terms which vary as $\rho$ on the pupil and vary linearly with the misalignments or less yields the expression for the distortion contribution of a single mirror which is despaced by $W$, decentered by $\vec \ell$ and tilted by $\vec \alpha$:\footnote{As noted in \S \ref{sec:generic_patterns}, we have omitted the $W_{311}$ coefficient as it is equal to zero.}

\begin{align}
\label{eq:Me_distortion}
G_i^{distortion} &= - \left[ 2\frac{W}{R}\left(\frac{W}{s} - 1\right)^2 \left( W_{220} +  W_{222} \right) + 3\frac{W^2}{R^2}\left(\frac{W}{s} - 1 \right)W_{131} + 4 \frac{W^3}{R^3} W_{040} \right] (\frac{\vec{\rho}}{R}\cdot\vec{\sigma})(\vec{\sigma}\cdot\vec{\sigma}) \\
\nonumber     & \qquad - \Bigg[ 4\left(\frac{W}{s}\right)\left(\frac{W}{s} - 1\right)W_{220} + 2\left(\frac{W}{s} -1 \right) \left(\frac{2W}{s} - 1\right)W_{222} \\
\nonumber     & \qquad \qquad \qquad \qquad \qquad \qquad  + 4\frac{W}{R}\left(\frac{3W}{2s} - 1\right)W_{131} + 8\frac{W^2}{R^2}W_{040} \Bigg] (\frac{\vec{\rho}}{R}\cdot\vec{\sigma})(\vec{\sigma}\cdot\frac{\vec{l}}{R}) \\
\nonumber     & \qquad - \left[ 2\frac{W}{R}\left(\frac{W}{s} -1 \right)\left(2 W_{220} + W_{222}\right) + 2\frac{W^2}{R^2}W_{131} \right] (\frac{\vec{\rho}}{R}\cdot\vec{\sigma})(\vec{\sigma}\cdot\vec{\alpha}) \\
\nonumber     & \qquad - \Bigg[2\left(\frac{W}{s} - 1\right)^2W_{220} + 2\left(\frac{W}{s}\right)\left(\frac{W}{s} - 1\right)W_{222}\\
\nonumber     & \qquad \qquad \qquad \qquad \qquad \qquad + 2\frac{W}{R}\left(\frac{3W}{2s} - 1\right)W_{131} + 4\frac{W^2}{R^2}W_{040} \Bigg] (\frac{\vec{\rho}}{R}\cdot\frac{\vec{l}}{R})(\vec{\sigma}\cdot\vec{\sigma}) \\
\nonumber     & \qquad - \left[ 2\frac{W}{R}\left(\frac{W}{s} -1 \right)W_{222} + \frac{W^2}{R^2}W_{131} \right] (\frac{\vec{\rho}}{R}\cdot\vec{\alpha})(\vec{\sigma}\cdot\vec{\sigma})
\end{align}

For a two mirror telescope, only the secondary contributes to the distortion wavefront delay, as the primary is by construct not despaced from its entrance pupil.  Grouping the coefficients in the above equation into terms of field dependences yields the form for distortion in \S \ref{sec:generic_patterns}:  

\begin{align}
\label{eq:section2_distortion}
G^{distortion} & = G^{distortion}_{Seidel}   \sigma^3 \rho \cos(\phi - \theta) \\
\nonumber             & + G^{distortion}_{decenter,\sigma} 
              \sigma^2 \ell \cos(\theta - \phi_\ell) \rho \cos(\phi - \theta)
              + G^{distortion}_{decenter,\rho} 
              \sigma^2 \ell \rho \cos(\phi - \phi_{\ell}) \\
\nonumber             & + G^{distortion}_{tilt,\sigma} 
              \sigma^2 \alpha \cos(\theta - \phi_\alpha)\rho \cos(\phi - \theta)
              + G^{distortion}_{tilt,\rho} 
              \sigma^2 \alpha \rho \cos(\phi - \phi_{\alpha})
\end{align}

\noindent The coefficients are given by:

\begin{align}
\label{eq:transition1_distortion}
G^{distortion}_{Seidel} &= \frac{1}{R} \left[ 2\frac{W}{R}\left(\frac{W}{s} - 1\right)^2 \left( W_{220} +  W_{222} \right) + 3\frac{W^2}{R^2}\left(\frac{W}{s} - 1 \right)W_{131} + 4 \frac{W^3}{R^3} W_{040} \right]\\
\nonumber G^{distortion}_{decenter,\sigma} &= -\frac{1}{R^2} \Bigg[ 4\left(\frac{W}{s}\right)\left(\frac{W}{s} - 1\right)W_{220} + 2\left(\frac{W}{s} -1 \right) \left(\frac{2W}{s} - 1\right)W_{222} \\
\nonumber              & \qquad \qquad + 4\frac{W}{R}\left(\frac{3W}{2s} - 1\right)W_{131} + 8\frac{W^2}{R^2}W_{040} \Bigg] \\
\nonumber G^{distortion}_{decenter,\rho} & = -\frac{1}{R^2} \Bigg[2\left(\frac{W}{s} - 1\right)^2W_{220} + 2\left(\frac{W}{s}\right)\left(\frac{W}{s} - 1\right)W_{222} \\
\nonumber              & \qquad \qquad + 2\frac{W}{R}\left(\frac{3W}{2s} - 1\right)W_{131} + 4\frac{W^2}{R^2}W_{040} \Bigg] \\
\nonumber G^{distortion}_{tilt,\sigma} & =  -\frac{1}{R} \left[ 2\frac{W}{R}\left(\frac{W}{s} -1 \right)\left(2 W_{220} + W_{222}\right) + 2\frac{W^2}{R^2}W_{131} \right] \\
\nonumber G^{distortion}_{tilt,\rho} & = -\frac{1}{R} \left[ 2\frac{W}{R}\left(\frac{W}{s} -1 \right)W_{222} + \frac{W^2}{R^2}W_{131} \right] 
\end{align}

\noindent where all mirror properties and object distances refer to the secondary mirror, and the chief ray angle is the chief ray angle on the primary mirror.

\end{document}